\documentclass[a4paper,onecolumn,superscriptaddress,11pt,accepted=2019-06-16]{quantumarticle}
	
\pdfoutput=1

\usepackage[english]{babel}
\usepackage[T1]{fontenc}
\usepackage[utf8]{inputenc}
\usepackage[
  backend=bibtex,%
  style=phys,%
  articletitle=false,%
  doi=false,%
  pageranges=false,%
  biblabel=brackets%
  ]{biblatex}
\bibliography{library.bib}
\DeclareFieldFormat{titlecase}{#1}

\usepackage{xparse}
\usepackage{amssymb}
\usepackage{amsmath}
\usepackage{mathrsfs}
\usepackage{amsthm}
\usepackage{mathdots}
\usepackage{physics}
\usepackage{paralist}
\usepackage{array}
\usepackage{graphicx}
\usepackage{subfigure}
\usepackage{wrapfig}
\usepackage[small]{caption}
\usepackage[inline,shortlabels]{enumitem}
\usepackage{calc}
\usepackage{listings}
\usepackage{xcolor}
\usepackage{hyperref}
\hypersetup{
	linktocpage,
    colorlinks = false 
    }

\DeclareMathOperator{\supp}{supp}
\DeclareMathOperator{\poly}{poly}
\DeclareMathOperator{\diam}{diam}
\DeclareMathOperator{\id}{id}
\newcommand{\iu}{\mathrm{i}}
\newcommand{\compl}[1]{{#1}^{c}}

\newcommand{\fastto}[1][]{\xrightarrow{#1}}
\newcommand{\fastfrom}[1][]{\xleftarrow{#1}}
\newcommand{\samephase}{\leftrightarrow}
\newcommand{\lX}{\bar{X}}
\newcommand{\lZ}{\bar{Z}}
\newcommand{\ldX}{\tilde X}
\newcommand{\ldZ}{\tilde Z}
\newcommand{\GS}[1]{\mathcal H_0(#1)} 
\newcommand{\ketbrasub}[3]{\ket{#1}_{#3}\!\bra{#2}}

\newcommand{\eps}{\varepsilon}
\newcommand{\err}[1]{O\left(#1\right)}
\NewDocumentCommand{\experr}{O{1} O{}}{%
\frac{#1}{\sqrt{2\pi}}\frac{e^{-\frac{1}{2}\left(\frac{\sqrt{T_{#2}}\eps}{2\tau_{#2}}\right)^2}}{\frac{\sqrt{T_{#2}}\eps}{2\tau_{#2}}}%
}
\NewDocumentCommand{\ceil}{s O{} m}{%
  \IfBooleanTF{#1} 
    {\left\lceil#3\right\rceil} 
    {#2\lceil#3#2\rceil} 
}
\newcommand{\fat}[2][]{\mathbb{F}_{#1}\big(#2\big)}

\newtheorem*{mainplain}{Main result}
\newenvironment{main}[1]{%
    \begin{mainplain}#1}{%
    \end{mainplain}%
	\setcounter{step}{0}
}
\newtheorem{step}{Step}
\newtheorem{substep}{Step}[step]
\newtheorem*{finalstep}{Final Step}
\newtheoremstyle{def name}
{}
{}
{\itshape}
{}
{\bfseries}
{.}
{ }
{\thmname{#1}\thmnumber{ #2}\thmnote{ (#3)}}
\theoremstyle{def name}
\newtheorem{definition}{Definition}

\newcounter{descriptcount}
\newlist{enumdescript}{description}{2}
\setlist[enumdescript]{%
  before={\setcounter{descriptcount}{0}%
          \renewcommand*\thedescriptcount{(\textit{\roman{descriptcount}})}}%
  ,font=\stepcounter{descriptcount}%
  \thedescriptcount~%
}
\makeatletter
\def\reflabel{
    \def\@currentlabel{\thedescriptcount}%
    \phantomsection%
    }
\makeatother

\begin{document}

\title{Classification of phases for mixed states via fast dissipative evolution}
\date{\today}
\author{Andrea Coser}
\affiliation{Departamento de An\'alisis Matem\'atico y Matem\'atica Aplicada, Universidad Complutense de Madrid, 28040 Madrid, Spain.}
\affiliation{ICMAT, C/ Nicol\'as Cabrera, Campus de Cantoblanco, 28049 Madrid, Spain.}
\author{David P\'erez-Garc\'ia}
\affiliation{Departamento de An\'alisis Matem\'atico y Matem\'atica Aplicada, Universidad Complutense de Madrid, 28040 Madrid, Spain.}
\affiliation{ICMAT, C/ Nicol\'as Cabrera, Campus de Cantoblanco, 28049 Madrid, Spain.}

\maketitle

\begin{abstract}
We propose the following definition of topological quantum phases valid for mixed states: two states are in the same phase if there exists a time independent, fast and local Lindbladian evolution driving one state into the other.
The underlying idea, motivated by~\cite{Konig2014}, is that it takes time to create new topological correlations, even with the use of dissipation. 

We show that it is a good definition in the following sense:
\begin{enumerate*}[label=(\arabic*)]
\item It divides the set of states into equivalent classes and it establishes a partial order between those according to their level of ``topological complexity''.
\item It provides a path between any two states belonging to the same phase where observables behave smoothly. 
\end{enumerate*}

We then focus on pure states to relate the new definition in this particular case with the usual definition for quantum phases of closed systems in terms of the existence of a gapped path of Hamiltonians connecting both states in the corresponding ground state path.
We show first that if two pure states are in the same phase in the Hamiltonian sense, they are also in the same phase in the Lindbladian sense considered here. 

We then turn to analyse the reverse implication, where we point out a very different behaviour in the case of symmetry protected topological (SPT) phases in 1D.
Whereas at the Hamiltonian level, phases are known to be classified with the second cohomology group of the symmetry group, we show that symmetry cannot give any protection in 1D in the Lindbladian sense: there is only one SPT phase in 1D independently of the symmetry group. 

We finish analysing the case of 2D topological quantum systems.
There we expect that different topological phases in the Hamiltonian sense remain different in the Lindbladian sense.
We show this formally only for the $\mathbb{Z}_n$ quantum double models $D(\mathbb{Z}_n)$.
Concretely, we prove that, if $m$ is a divisor of $n$, there cannot exist any fast local Lindbladian connecting a ground state of $D(\mathbb{Z}_m)$ with one of  $D(\mathbb{Z}_n)$, making rigorous the initial intuition that it takes long time to create those correlations present in the $\mathbb{Z}_n$ case that do not exist in the $\mathbb{Z}_m$ case and that, hence, the $\mathbb{Z}_n$ phase is strictly more complex in the Lindbladian case than the $\mathbb{Z}_m$ phase.
We conjecture that such Lindbladian does exist in the opposite direction since Lindbladians can destroy correlations.
\end{abstract}

\newpage

\tableofcontents

\section{Introduction}

Motivated by the every-day increasing experimental capabilities to engineer exotic topological phases~\cite{Sau2010,Xiao2011,Goldman2016a}, and by their potential application to quantum technological tasks~\cite{Nayak2008,Hasan2010,Qi2011}, there has been an important collective effort in the last years devoted to understand and classify {\it all} possible topological phases that can appear in quantum many body systems~\cite{Wen2017b}.

Intuitively the concept of phases is strictly connected to analiticity:
while going from one phase to another we expect something to happen (criticality).
For example imagine a system described by a one-parameter Hamiltonian.
By tuning the parameter, if we go through a phase transition, we expect to see some critical behaviour in the ground state, like the decay of correlations going from an exponential law to a power law.
Of course such critical behaviour cannot strictly happen at finite system size $N$, where everything is analytic.
Even if we can see signs of a phase transition at finite size, in order to see true critical behaviour the thermodynamic limit $N\to\infty$ is fundamental, since only in this limit analyticity may be lost.

A mathematical definition of gapped quantum phases is available for the case of pure states through paths of gapped Hamiltonians.
Two models described by two local gapped Hamiltonians $H_0$ and $H_1$ are in the same phase if there exists a path $H(s)$ for $s\in[0,1]$ of local Hamiltonians such that $H(0)=H_0$, $H(1)=H_1$ and $H(s)$ is gapped uniformly in $s$, namely the gap $\lambda(s)$ between the ground state space and the first excited level of $H(s)$, is uniformly bounded by a constant in the thermodynamic limit, $\lambda(s)\geq \lambda > 0$.
Notice that if $H(s)$ is gapped, then $\expval{O}_s$, the expectation value of any local operator $O$ on the ground state of $H(s)$, is smooth in $s$.
Indeed it is possible to prove that if such a gapped path of Hamiltonian exists, then the final ground state can be obtained by unitary evolving the initial one, with the so called quasi-adiabatic Hamiltonian.
Any expectation value $\expval{O}_s$ can be obtained as the evolved of $\expval{O}$ for a finite time $s$.
The quasi-adiabatic Hamiltonian is (quasi-)local, and since any local Hamiltonian evolution provides a well-defined limit in the thermodynamic limit~{\cite{Nachtergaele2006a}}, $\expval{O}_s$ cannot show any non-analyticity (see e.g.~\cite[Theorem 6.2.4]{Bratteli2012} for a mathematical proof of this 
fact).

In practice systems are never isolated and noise may play an important role while describing and analysing topological phases.
It is known for instance that in 2D some topological properties vanish in the thermodynamic limit {\it at any} non-zero temperature~\cite{Castelnovo2007,Castelnovo2008,Iblisdir2009,Hastings2011}.
This problem can be inserted in the wider problem of defining phases for generic quantum mixed states.
In this case there may be no Hamiltonian whatsoever, so it makes more sense to consider phases as classes of equivalence of states, instead of models.

For pure states, these two points of view are equivalent.
Indeed, by definition two pure states $\ket{\phi_0}$ and $\ket{\phi_1}$ are in the same phase if there exist two Hamiltonians $H_0$ and $H_1$ such that $\ket{\phi_0}$ and $\ket{\phi_1}$ are their respective ground states and such that they can be connected by a gapped path $H(s)$.
Notice also that given such path, any pair of gapped Hamiltonians $H_0'$ and $H_1'$ which have $\ket{\phi_0}$ and $\ket{\phi_1}$ as ground states are in the same phase.

In this work we give a tentative definition of phases for mixed states.
The main ingredient of our definition is the dissipative Markovian evolution described by a Lindbladian superoperator.
Loosely speaking, two generic states are in the same phase if we can devise such an evolution to drive in short times one state to the other.
We will see that our definition will also provide a partial order in the set of phases.
This picture of a more complex structure of the phase diagram of mixed states has first been proposed in Ref.~{\cite{Konig2014}}.
In this work we will follow on the lines set out therein, by giving a formal definition of phases through Lindbladian dissipative dynamics.
We will show that it is a good definition in the sense that it defines classes of equivalence, and it connects nicely with the topological content of a state and with analyticity of observables.
We will investigate many connections with the usual Hamiltonian definition for pure states.
With respect to that, the dissipative definition also gives a partial order in the set of equivalence classes, thus providing a more sophisticated structure in the set of equivalence classes.
On top of that it has a clear operational meaning.
We will formalize the definition in Section~\ref{subsec:formal def}.

Our definition is not the first attempt to define phases (and in particular topological phases) in the context fo open quantum systems.
An important contribution was given by Diehl et al.~\cite{Diehl2011,Bardyn2012,Bardyn2013} for Guassian states.
Their ideas sparkled a large amount of subsequent work (for example see~\cite{Rivas2013,Viyuela2014}).
In particular, a generalisation was proposed in~\cite{Grusdt2016}.
However, we take here a different point of view, and the exact connections between their definition and ours is an open question.
We will give more details in Section~\ref{subsec:previous work}.

The point of view of phases of quantum states, as opposed to phases of Hamiltonians, is common in quantum information, where two states are considered to be in the same phase if and only if there exists a unitary and local finite depth quantum circuit that transforms one state into the other.
It is generally believed that two pure states can be connected by a gapped path of Hamiltonians if and only if there exists a local finite depth quantum circuit that transforms one into the other.
Any rigorous proof of this general statement must involve non trivial approximation analysis depending on the specific purpose.
In Section~{\ref{sec:pure case}} we will need an exact result 
as an intermediate step to connect our proposed definition to the Hamiltonian ones for pure states.
Given the quasi-adiabatic Hamiltonian evolution, we will need to approximate it with a finite depth quantum circuit and obtain some precise bounds on the errors for our purposes later on.
Some results along this line can be found in Refs.~\cite{Osborne2007,Huang2015}, however they are not directly applicable to our case.
Indeed, they connect the initial and final local density matrices, while we are interested in the full states.
The specific result we need can be obtained following the same lines of the proofs in Refs.~\cite{Osborne2005,Haah2018}, generalizing them for the case of the quasi-adiabatic evolution.
For completeness we will give a rigorous proof of the result we need in Section~\ref{subsec:finite depth circuit}.
In particular, we will show that if two states are connected by a path of gapped Hamiltonians, then they are related by a finite depth unitary quantum circuit with gates that act on regions whose size is $\poly$-logarithmic in the system size.
Such a quantum circuit is not strictly local, however the result is enough for our scopes.

The definition through quantum circuits seems more suitable to be extended to the case of mixed states, where one can simply allow for noisy gates; i.e.\ gates that are not necessarily unitaries but completely positive trace preserving (CPTP) maps. The continuous time version of such CPTP circuits are semigroups of CPTP maps whose generators, named Lindbladians, have a local structure. In this context, the analogous of a finite depth quantum circuit is a fast dissipative Lindbladian evolution.
Here fast means that the evolved state is $\eps$-close to the final one in a time which is less than linear in the linear dimension of the system, for example ($\poly$-)logarithmic in the system size, with $\eps$ vanishing in the thermodynamic limit.
Notice that, as discussed at the beginning of this Introduction, the concept of phases is strictly speaking well defined only in the thermodynamic limit.

Let us now draw an intuitive but hand wavy picture, in order to explain what is the idea behind the definition of phase that we want to propose in this work.
Loosely speaking we say that two generic mixed states are in the same phase if they can be driven one to the other with fast dissipative evolution (in the sense explained above).
Locality and fast evolution will guarantee that new kinds of long range correlations, other than the ones already present in the initial state, cannot be created.
Hence the two states must contain the same kind of long range correlations and in this sense, they are in the same phase.

As already anticipated, this definition also provides a partial order on the set of equivalence classes.
Indeed, given two states $\rho_1$ and $\rho_2$, it could be that the long range correlations of $\rho_1$ are a subset of the ones of $\rho_2$.
Therefore we expect that no dissipative evolution could bring $\rho_1$ to $\rho_2$ fast, since it needs to create new long range correlations.
On the contrary we could expect that we can drive $\rho_2$ to $\rho_1$ fast, since we expect that destroying certain correlations can be done with local evolution in short times.
An example of this phenomenon has been studied in Ref.~\cite{Konig2014}.
While any state can be driven to the product state with a very fast Lindbladian, the authors show that there are states that cannot be created from the product state with dissipative evolution in a time shorter than the linear system size.
We conjecture here something stronger.
We expect there exist states which are not in the trivial phase and which can be reached with a fast evolution from other states in non trivial phases.
At the same time such fast evolution cannot be reversed:
No Lindbladian evolution would prepare in short time the latter state starting from the former.

Let us notice also that recent works have shown how the mixing time of a Lindbladian emerges as an important element, more than others like the spectral gap~\cite{Szehr2015a}, while analyzing its long term behaviour%
\footnote{Notice that a finite gap does not imply fast mixing time.
Even at the classical level there are known cases of Lindbladians with a uniform finite gap, but with mixing time that scales linearly in the system size~\cite{Aldous2001,Ganguly2015}.
In both classical and quantum systems the appearance of cutoff phenomena~\cite{Kastoryano2012,Kastoryano2013} is one feature, other than the gap, which may govern the convergence of the dissipative evolution.}
(e.g.\ stability in observable quantities~\cite{Cubitt2015}, or area-laws in the fixed points~\cite{Brandao2015}).
However, here we are not strictly talking about the mixing time of the Lindbladian.
We request the evolution to be fast only between the two given states, not starting from any state.

The first question one may ask is what happens when we restrict this new definition of phases to the well studied case of pure states.
Is the resulting classification of phases the same as the usual via the path of gapped Hamiltonians?
We will partially answer this question in Section~\ref{sec:pure case} by showing that if two pure states are in the same phase according to the standard Hamiltonian classification, then they are connected by a fast Lindbladian evolution.

In Section~\ref{sec:other implication} we will comment and give some examples of the other implication.
In Section~\ref{subsec:1D SPT phase} we will see that the Lindbladian definition is not completely equivalent to the Hamiltonian one, at least in the case of one dimensional systems with symmetry restrictions.
We will show indeed that the dissipative definition of phases gives a totally different classification for 1D symmetry protected topological (SPT) phases.
In particular, it gives a unique SPT phase in 1D.

In Section~\ref{subsec:slow Zm to Zn}, we show that in two dimensions we cannot obtain certain topologically ordered states as the result of a fast dissipative evolution if the topological content of the initial state is a `subset' of the one of the final state%
\footnote{Here the precise meaning of the topological content of a state being `included' in the topological content of another state is intentionally left vague.
What we exactly mean with this statement should be clarified in Section~\ref{sec:other implication}.}.
The results of Sections~\ref{subsec:1D SPT phase} and~\ref{subsec:slow Zm to Zn} show that somehow the symmetry protection in 1D is weaker than the topological protection in 2D since the former can be overcome with a symmetry preserving Lindbladian evolution, while the latter cannot.

In Section~\ref{subsec:anyon condensation} we will also discuss the possibility of finding some fast evolution that connects a state with some topological order to another state which contains a subset of the initial topological order.
As an example, we will consider the case in which the topological order in the initial state is described by some group $G$, and in the final state it is described by a group $H$ which is a normal subgroup of $G$.
We will give an explicit example of this phenomenon for GHZ states in one dimension.
Even if we are not able to give examples for topological states in two dimensions, we conjecture that it is related to the mechanism of anyon condensation.

\section{Definition of phases for mixed states}
\label{sec:definition}

Consider a many-body spin system living in a lattice $\Lambda$, with Hilbert space $\mathcal H_S = \bigotimes_{i\in\Lambda}\mathcal H_{S(i)}$.
Let $N$ be the number of sites in $\Lambda$, and $L$ the maximal linear size of $\Lambda$ (its diameter).
Loosely speaking, we say that a state $\rho_0$ can be driven fast to another state $\rho_1$, both supported on $\mathcal H_S$, and we write $\rho_0\fastto\rho_1$, if it exists a dissipative evolution generated by a local and time-independent Lindbladian $\mathcal L$ such that  for $t\gtrsim \poly\log N$,
\begin{equation}
\label{eq:informal definition}
\norm{e^{t\mathcal L}(\rho_0) - \rho_1}_1 \lesssim \poly(N)\, e^{-\mu t} \,.
\end{equation}
Notice that after a time $t\gtrsim \poly\log N$, the upper bound will vanish in the thermodynamic limit.
For this reason, the time scale $\mu^{-1}$ can be chosen constant or of order $\log N$.

The relation defined in eq.~\eqref{eq:informal definition} allows us to give a definition of phases for Lindbladians.
We say that two states belong to the same phase if there exist two local Lindbladian evolutions such that $\rho_0 \fastto \rho_1$ and $\rho_0\fastfrom\rho_1$, and in this case we write $\rho_0 \samephase \rho_1$.

\subsection{The formal definition}
\label{subsec:formal def}

Despite the simplicity of the definition in eq.~\eqref{eq:informal definition}, there are many subtleties that one has to take into account.
We make here a series of observations and in light of them, we will give a more formal and complete definition of phases at the end of this section.

\begin{enumdescript}[wide,labelindent=0pt]
\item[The thermodynamic limit.]
\reflabel\label{it:thermodynamic limit}
First of all, it is important to remark that in our definition of phases we are interested in how the system behaves in the thermodynamic limit $N\to\infty$.
It is well known that phase transitions may appear only when we take this limit.
Indeed, for finite system size $N$ everything is smooth, no singularities can appear, and every state is in the same phase.
The thermodynamic limit is therefore an essential tool to study phases and phase transitions.
However, from a mathematical point of view, actually taking the thermodynamic limit leads in general to many complications and a formal treatment is usually very involved.
Here we are not interested in the formalities of the thermodynamic limit, hence we take a different, more physics-oriented approach:
We will only compute how quantities and error estimates scale with the system size, without ever formally taking the limit $N\to\infty$.

Even if we never take the formal limit, it is worth remarking that we are implicitly considering systems were the thermodynamic limit makes sense.
In particular we consider systems which can be unambiguously defined for different system sizes, as for example systems with a certain degree of translational symmetry.

\item[No need of fast mixing.]
\reflabel\label{it:fast mixing}
We don't require the Lindbladian $\mathcal L$ to be fast mixing.
The mixing time for the evolution of some state other than $\rho_0$ may be slow.
We require it to be fast only when applied to $\rho_0$.

\item[Locality and the Lieb-Robinson bounds.]
\reflabel\label{it:LR bounds}
In the definition we required the Lindbladian to be local, namely it can be written as a sum of local terms
\begin{equation}
\label{eq:local L}
\mathcal L = \sum_{X\subset \Lambda} \mathcal L_X , \qquad \norm{\mathcal L_X} \sim O(1), \qquad \supp[\mathcal L_X] = X, \qquad |X|<k,
\end{equation}
for some fixed $k$.
However, according to the specific needs, one can relax this assumptions and ask for quasi-locality in the sense of \cite{Bachmann2012}, or allow for the size of $X$ to be polylogarithmic in the system size $N$, while keeping the norm of every single term of order one.

The need for some notion of locality is in order to avoid correlations to spread too fast.
Lieb-Robinson bounds for local evolutions give an upper limit to the speed at which correlations can be created throughout the system.
Therefore it makes sense to consider the time needed to drive a state into another one.
If in the final state there are long range correlations that are not present in the initial state, the Lieb-Robinson bounds will provide a lower bound to the time needed to drive the latter state to the former one.

Allowing for non local evolution would instead make the definition trivial.
For any states, eq.~\eqref{eq:informal definition} could always be satisfied with a non local Lindbladian.
A highly non local evolution could indeed generate in short times correlations at a distance comparable with the system size $N$, thus connecting for example states with different long range (topological) order.
The generated phase diagram would then have a single phase.

In general we require the time needed to generate correlations at a distance of the order of the linear size $L$ of system to be $\sim L$, apart from some possible logarithmic terms.
For example, in Section~\ref{sec:pure case} we will need to consider Lindbladians of the form of eq.~\eqref{eq:local L} but with $\diam(X)\lesssim \log^\alpha L$.
In this case, we can think of a new lattice $\Lambda'$ obtained by blocking $(\log^\alpha L)^d$ sites of the original one. 
The Lindbladian in $\Lambda'$ is local with terms supported on a finite number of blocked sites.
The new system has linear dimension $L'\sim L/\log^\alpha L$, and Lieb-Robinson bounds tell us that after a time $t$, correlations at a distance $x'$ are bounded by $\sim e^{vt - ax'}$.
Here $x'$ is the distance between two blocked sites measured in units of the unit cell of $\Lambda'$.
We can rewrite $x'$ in terms of $x$, the distance between two sites in the original lattice $\Lambda$, as $x'\sim x/\log^\alpha L$.
Therefore, in the original lattice, correlations after a time $t$ at a distance $x$ are bounded by $\sim e^{vt - ax/\log^\alpha L}$.
This means that to generate long range correlation at a distance $x\sim L$, we need a time $t\sim L/\log^\alpha L$.

\item[Time independence of the Lindbladian.]
\reflabel\label{it:L time-indep}
We ask for the Lindbladian to be time-independent because the definition gets a nice operational interpretation:
For two states, we say that $\rho_0\fastto{\rho_1}$ if we can device an external source of noise and let the system flow quickly from $\rho_0$ to $\rho_1$ without acting on it any more.

\item[Adding ancillas to the definition.]
\reflabel\label{it:ancillas}
The definition~\eqref{eq:informal definition} should be extended to allow for some catalizer ancillas to be added to the initial state of the system $\rho_0$, typically in a product state.
Then the time-independent Lindbladian $\mathcal L$ drives the whole system with ancillas to the final state, where the ancillas are again decoupled from (in a product state with) the initial system and can be easily discarded.

For everything to make sense, we need some locality requirements on the interactions of the ancilla system.
We can clarify this point with a simple example:
Consider a finite set of ancilla qubits interacting among themselves, and also coupled to sites of the original system which belong to regions far apart from each other.
This interaction could clearly induce some non-locality in the effective action of the Lindbladian on the system, thus possibly generating long range correlations in very short times.
More concretely, if $\mathcal H = \mathcal H_S \otimes \mathcal H_T$, with $\mathcal H_S$ the Hilbert space of the original system and $\mathcal H_T$ the Hilbert space of the ancillas, the Lindbladian can be written as $\mathcal L = \mathcal L_S + \mathcal L_T + \mathcal L_{S,T}$, where $\mathcal L_S$ acts on the original system in $\mathcal H_S$ and it is local according to observation~\ref{it:LR bounds}, $\mathcal L_T$ acts only on $\mathcal H_T$ while $\mathcal L_{S,T}$ describes the interaction between the system and the ancillas.
In order for the ancillas not to destroy locality in $\mathcal H_S$ we first need to introduce a spatial structure for the ancilla qubits that form $\mathcal H_T$.
Any ancilla can then interact only with nearby sites.
For the purpose of this work, we need a set of ancilla qubits per lattice site, $\mathcal H_T=\bigotimes_{i\in\Lambda}\mathcal H_{T(i)}$.
Ancillary sistems corresponding to different lattice sites are independent from each others, $\mathcal L_T = \sum_{i\in\Lambda} \mathcal L_{T(i)}$.
Moreover, the ancillas in $\mathcal H_{T(i)}$ may interact only with the sites in the sets $X\subset\Lambda$ in eq.~\eqref{eq:local L} which contain the site $i$.
More precisely, $\mathcal L_{T,S} = \sum_{X\subset \Lambda} \sum_{i\in X} \mathcal L_{X,T(i)}$, with $\mathcal L_{X,T(i)}$ acting on $\left(\bigotimes_{j\in X}\mathcal H_{S(j)}\right)\otimes \mathcal H_{T(i)}$.
Notice that another way to destroy locality would be to allow for a very strong interaction between the original system and the ancillas.
To avoid that, we also need the norm of the local terms $\mathcal L_{X,T(i)}$ to be of the same order of the local terms acting only on the original system, $\mathcal L_X$.
Otherwise, Lieb Robinson bounds could be violated.
On the contrary, the interaction strength in the ancillary system does not contribute to the Lieb-Robinson bounds in the original system.
Therefore, we do not need any special requirements on $\norm{\mathcal L_T}_{1\to 1}$.
These requirements are enough for the purposes of our work, however a different description of the ancillas may be used for some other tasks.
The important point here is that a `good' ancillary system should not introduce any non-local effective interactions on the original system.

One may also want to ask the size of this ancillary system not to be too large.
This requirement does not seems as fundamental as the locality preserving one, since even a huge size of the ancilla space would not influence Lieb-Robinson bounds in the system.
However, if one wants a Lindbladian that could possibly be devised in a lab, at least in principle, some bounds on the growth of the size of the ancillas with the system size are necessary.
It seems reasonable to ask for the ancillary system attached to each site to have at most $\poly N$ qubits, so that $\dim\mathcal H_{T(i)} \sim 2^{N^\beta}$ for some $\beta>1$.
In this case it also seems reasonable to ask for $\mathcal L_{T(i)}$ to have the same degree of locality as the original system, in the sense that the interaction terms must involve only a finite number of ancilla qubits (or logarithmic, in case the original interaction was $\log$-local).
While these seem reasonable requirements for the size of the ancilla space, for the purposes of this work we can ask something stricter.
The Hilbert space of the ancillas that we are going to use in Section~\ref{subsec:transitivity} has indeed dimension exponential in $\poly N$, however the states that are actually accessible for the particular dynamics that we are going to use are of order $\sim N^\beta$.
This means that we could consider another description of the ancillas in an Hilbert space $\mathcal H_{T(i)}'$ with dimension $\dim\mathcal H_{T(i)}'\sim N^\beta$, that can be described with $\sim\beta \log_2 N$ qubits.
In this case the locality in $\mathcal H'_{T(i)}$ is lost and the interaction terms in $\mathcal L_{T(i)}$ could involve all such qubits.
We are trading a logarithmic term in the locality with a smaller size of the ancillary system.

We call \emph{locality preserving} any set of ancillas with the above properties for their size and the locality of the interactions among themselves and with the original system.

As a final remark, we point out that the requirement for the ancillas to be in a product state at the end of the evolution may be too strong.
We could also ask for the ancillas to be simply traced out in the end, however having the ancillas in a product state gives a nicer operational interpretation, since they can be discarded deterministically.

\item[States along the fast evolution are in the same phase.]
\reflabel\label{it:intermediate states}
For our definition of phases to actually define an equivalence relation, and an order on the set of the equivalence classes, we need it to be transitive.
If we require the Lindbladian $\mathcal L$ to be time independent, then transitivity is not manifest in the definition.
However, it turns out to be true, and we show it in Section~\ref{subsec:transitivity}.
With the results and the techniques of that section, we can also see that given $\rho_0\fastto[\mathcal L]\rho_1$, we have that for $\rho_{0,t}=e^{t\mathcal L}(\rho_0)$, $\rho_{0,t}\fastto[\mathcal L]\rho_1$ and $\rho_0\fastto[\mathcal L']\rho_{0,t}$.
While the former statement is obvious, for the latter we need to construct the Lindbladian $\mathcal L'$, by coupling $\mathcal L$ to a timer (as described in Section~\ref{subsec:transitivity}) such that it turns off the evolution after a time $t$.
Moreover, if $\rho_0\samephase \rho_1$ then by transitivity and using this last observation, every state $\rho_{0,t}$ is in the same phase of $\rho_0$ and $\rho_1$.
The same is valid for any state $\rho_{1,t}$ obtained evolving $\rho_1$.

\item[Analiticity propeties of the fast dynamics.]
\reflabel\label{it:smooth evolution}
Another observation related to the previous point concerns the nature of the states along the fast evolution.
In the classical description of phases and phase transitions, we expect that if we stay within the same phase, the physical properties of the system will change smoothly.
On the contrary, if we cross a phase transition, a global change must happen in the state.
For standard phases, one can find some local order parameters which show non-analytic behaviour when crossing a phase transition.
Of course for topological phases this is not true, and we can actually change phase while all local observable change smoothly.

We have seen in observation~\ref{it:intermediate states} that if $\rho_0\samephase \rho_1$, then all states along the two evolutions from $\rho_0$ to $\rho_1$ and from $\rho_1$ to $\rho_0$, are in the same phase.
Therefore, we expect that along these two evolutions all local observables behave smoothly.
Indeed, notice that for any finite $t$, the evolution $\Tr_{\compl{A}}[e^{t\mathcal L}(\rho_0)]$ restricted to a region $A$ of finite size is analytic in $t$%
\footnote{The proof is totally analogous to the Hamiltonian case and can be done easily following the steps of~\cite[Theorem 6.2.4]{Bratteli2012}, see also~\cite{Nachtergaele2011a}.}.
However, our definition eq.~\eqref{eq:informal definition}, requires times of order $t\sim\poly\log N$ to reach the final state.
In the limit $N\to\infty$ the time goes to infinity, even if very slowly, and this could lead to non-analyticities.
We can overcome this problem by using the results of local fast mixing of Ref.~\cite{Lucia2015,Cubitt2015}.
While they study the case of fast mixing from any initial state, their result can be translated to the present setting when the evolution is fast only from a specific initial state $\rho_0$.
In this case, their result reads
\begin{equation}
\norm{\Tr_{\compl{A}}[e^{t\mathcal L}(\rho_0)-\rho_1]}_1 \lesssim \poly(|A|)\, e^{-\mu t}.
\end{equation}
The main difference with eq.~\eqref{eq:informal definition} is in the prefactor of the exponential decay.
According to Ref.~\cite{Lucia2015,Cubitt2015}, for local observables this prefactor does not depend on the total system size, but only on the size of the support of the observable.
Therefore, after a possibly large but finite time (independent of $N$), the evolution of such observable will have reached its final value.
For what discussed before, along this very short evolution, any such local observable cannot show any non-analyticity.

\end{enumdescript}

In light of these observations, we are now ready to give a formal definition of phases through Lindbladian evolution.

\begin{definition}[Lindbladian phases]
\label{def:L definition}
We say that a state $\rho_0$ can be driven fast to another state $\rho_1$, and we write $\rho_0\fastto\rho_1$, if there exists a dissipative evolution generated by a local (according to observation~\ref{it:LR bounds}) and time-independent Lindbladian $\mathcal L$ acting on the original system and a \emph{locality preserving} ancillary system (as defined in observation~\ref{it:ancillas}), such that for any $t\gtrsim \poly\log N$
\begin{equation}
\label{eq:formal definition}
\norm{e^{t\mathcal L}(\rho_0\otimes \sigma_0) - \rho_1\otimes \sigma_1}_1 \lesssim \eps_N ,
\end{equation}
with $\sigma_0$ and $\sigma_1$ respectively the initial and final states of the ancillas, and $\eps_N\rightarrow 0$ in the thermodynamic limit $N\rightarrow\infty$.

We say that two states belong to the same phase if there exist two local Lindbladian evolutions as described above such that $\rho_0 \fastto \rho_1$ and $\rho_0\fastfrom\rho_1$, and in this case we write $\rho_0 \samephase \rho_1$.
\end{definition}

\subsection{Connections with previous work}
\label{subsec:previous work}

In this section, we comment on previous works which are related to our definition of phases.
There is a wide literature about how to characterize topological order in the case of mixed states and in particular in the presence of a non vanishing temperature.
A natural question for ground states of topological models is whether the topological properties will survive when the system is put in contact with a thermal bath~\cite{Castelnovo2007,Castelnovo2008,Iblisdir2009,Hastings2011,Viyuela2011,Viyuela2012}.
This works are mainly focused on detecting some residual presence of topological order, rather than giving a full characterization of the phase these mixed states belong to.
Here we will focus mainly on works that make an attempt to give a (partial) definition of phases for open quantum systems.

As mentioned in the Introduction, in a series of works~\cite{Diehl2011,Bardyn2012,Bardyn2013} Diehl et al.\ propose a way for classifying phases for Gaussian mixed states.
Their approach is different in spirit from ours, since they try to generalize to the case of mixed states the usual definition of phase for pure states via the gapped path of Hamiltonians (see Definition~\ref{def:H definition}).
They consider fixed points of gapped Lindbladians, and they study what happens by smoothly deforming the state (or equivalently the corresponding Lindbladian).
They restrict themselves to fermionic systems for which the Lindblad operators are linear in the fermionic creation and annihilation operators.
This guarantees that the fixed point is Gaussian and therefore all the information is contained in the two-point equal time correlation matrix.
From the correlation matrix, the authors build a fictitious quadratic Hamiltonian from which they construct a topological classification of Gaussian states using the usual Hamiltonian scheme~\cite{Schnyder2008}, and identify some topological invariants.
A key difference with the Hamiltonian case is that two independent gaps have to be taken into account in the mixed case.
A \emph{purity gap}, namely the gap of the fictitious Hamiltonian built from the correlation matrix, and a \emph{dissipative gap}, namely the gap of the Lindbladian whose fixed state is the state under consideration.
The authors argue that the purity gap is the one responsible of determining the topological phase.
If it closes when smoothly deforming the state, the topological invariants may change abruptly and the state undergoes a phase transition.
On the other side they argue that the dissipative gap dictates whether the phase transition will be characterized by critical behaviour or not.
The authors provide examples of topological phase transitions where the dissipative gap stays open and the correlations remain exponentially decaying and others where the dissipative gap closes and the correlations decay polynomially. 

Notice that our approach takes a different but complementary point of view.
Instead of generalising the notion of phases via the gapped path of Hamiltonian and the smooth deformation of the state, we take a dynamical perspective.
In the Hamiltonian case the two points of view are equivalent, since the existence of a gapped path between two pure states guarantees, through the quasi-adiabatic theorem~\cite{Hastings2003}, the existence of a unitary local dynamics that connects the two states (see Section~\ref{subsec:finite depth circuit}).
Our definition tries to generalise this dynamical point of view, by requiring the existence of a fast dissipative dynamics connecting two mixed states in the same phase.
It is still an open question whether our definition coincides with the one of Diehl et al.\ in the case of Gaussian states.
Unfortunately, one big limitation of the approach by Diehl et al.\ is that it can only work for Gaussian states, since there is no clear generalisation of the concept of purity gap to general mixed states.

A step in this direction has been made in Ref.~\cite{Grusdt2016}, where the author considers quasi-thermal states, namely mixed states whose logarithm is a local Hamiltonian.
Similarly to the Gaussian case he can identify some kind of purity gap, as the gap of such local Hamiltonian.
In the case of a closed system, with the dynamics governed by a local Hamiltonian, the author defines states to be in the same phases iff they are connected by a local unitary transformation and studies the robustness of this definition under weak and strong local perturbations.
He also considers the case of open systems whose dynamics are governed by local Lindbladians.
Also in this case he defines phases through local unitary transformations.
In an appendix, he comments on the possibility of defining phases via a non unitary evolution, in a similar fashion to what we do in this work.
However, he concludes that this approach is unsuccessful because it would generate a single trivial phase.
This seems in contrast with what we find here, and the reason is that in~\cite{Grusdt2016} the author considers only fixed points of globally fast Lindbladians. 
In such case of course all states could be obtained very fast from the product state and one obtains a single trivial phase.
In this work instead we allow for Lindbladians whose time of convergence from the initial to the final state may scale with the system size.
As commented in detail in this section, we define two states to be in the same phase if such scaling is less than linear (say poly-logarithmic).
As already mentioned, this allows for a very rich structure of the phase diagram.

To conclude, let us comment on another topic that resonates with what we study in this work, but differs actually in several fundamental aspects, namely dynamical quantum phase transitions (DQPT).
For a review on the subject see~\cite{Heyl2018} and references therein.
This phenomenon refers to the emergence of nonanalyticities in the time evolution of certain observables after a quantum quench.
In this scenario the system is prepared in the ground state $\ket{\psi_0}$ of an initial Hamiltonian $H(\lambda_0)$ at a value $\lambda_0$ of some tunable parameter.
At time $t=0$ the parameter is suddenly switched to $\lambda_1$ and the state evolves according to $H(\lambda_1)$.
If $\ket{\psi_0}$ is not an eigenstate of $H(\lambda_1)$ the evolution is non trivial.
The typical quantity that is studied in this setting is the Loschmidt echo, which is the modulus square of the overlap of the evolved state with the original one, $\braket{\psi_0}{\psi_0(t)}=\expval{e^{-\iu H(\lambda_1)t}}{\psi_0}$.
DQPTs appear as nonanalyticities in the Loschmidt echo as a function of time, and can be thought as {\it phase transitions in time}.
They are closely connected with the usual equilibrium quantum phase transitions, since they often appear when the parameter of the Hamiltonian is quenched across an underlying equilibrium phase transition.
There are however examples in which no DQPT shows up even in the presence of an equilibrium QPT.

It could seem that the existence of DQPTs is in contradiction with our statement in observation~\ref{it:smooth evolution} in Section~\ref{subsec:formal def}, about the analyticity for all finite times of the Lindbladian (and therefore also the Hamiltonian) evolution.
This is actually not the case, and the reason stems from the fact that, in spite of some similarities between DQPTs and our definition of phases, there are some substantial differences.
First of all the result of analyticity of observation~\ref{it:smooth evolution} is valid for expectation values of local observables, while the Loschmidt echo is a highly non-local object (it can be seen as the expectation value of the projector onto the ground state of the initial Hamiltonian), therefore nonanalyticity may appear.
Moreover, while the Loschmidt echo probes the low energy (according to $H(\lambda_0)$) properties of $\ket{\psi_0(t)}$, the quench pumps energy into the system, and therefore the evolution of local observables are not dictated by the low energy physics only.
The nonanalyticities in the Loschmidt echo are in general smoothened out in the evolution of local observables due to the contribution of the high energy components of $\ket{\psi_0(t)}$.

Another important difference is that in our definition we fix an initial and final state and we look for a possible fast dissipative dynamics that connects the two.
If such dynamics exists, then the local observables will behave smoothly.
If instead the two states are in different phases, then no fast dynamics exist.
In this sense, in our context nonanalyticities never appear, because we never cross the phase transition dynamically.
Notice also that in the quantum quench case there is no final state, since the system is closed and the evolution unitary.
The expectation values of local observables may reach a steady state value (if the energy is the only conserved quantity they thermalize), but the full global state will not have a well defined infinite time limit.
This draws another fundamental difference with our definition of phases.

\subsection{Transitivity}
\label{subsec:transitivity}

In this section we show that the definition of phases through fast dissipative Markovian evolution actually divides the set of states in equivalence classes and provides a partial order among these classes.

The definition is clearly reflexive and symmetric, however if we do not allow for time dependent Lindbladians it is not manifestly transitive.
Let us consider three states, $\rho_a$ for $a=0,1,2$, with $\rho_0\fastto[\mathcal L_1]\rho_1$ and $\rho_1\fastto[\mathcal L_2]\rho_2$.
The Lindbladians $\mathcal L_1$ and $\mathcal L_2$ on the arrows are the ones that implement the fast evolution in eq.~\eqref{eq:formal definition}.
To show transitivity we need to find a Lindbladian $\mathcal L$ such that $\rho_0\fastto[\mathcal L]\rho_2$.
This is non trivial since we ask for the Lindbladian $\mathcal L$ to be time independent.
The strategy to build such an evolution is to add a \emph{locality preserving} ancillary system with the characteristics described in observation~\ref{it:ancillas} in Section~\ref{subsec:formal def}, whose internal evolution will provide a set of timers of the kind discussed in Ref.~\cite{Kastoryano2013}.
Some of these ancillas will act as a control on the local terms of the Lindbladian, changing from $\mathcal L_1$ to $\mathcal L_2$, and we call them switches.
The timers must be devised to turn all the switches at the same time and within a narrow time window.
In the final state, the ancillary system will be in a product state with respect ot the original system and can be discarded.
Moreover, if during the evolution the ancillary system is traced out, the effective evolution on the original system is a time dependent Lindbladian
$\mathcal L(s) = \theta(\tau-s)\mathcal L_1 + \theta(s-\tau)\mathcal L_2$.
The parameters of the timers has to be chosen such that the emerging time scale $\tau$ is short (say at most $\poly\log N$), but large enough so that the state 
$e^{\tau\mathcal L_1}(\rho_0)$ is close enough to $\rho_1$.
In the following we will detail the procedure in a more rigorous way.

\paragraph{Setup.}
We begin with the system living on a Hilbert space $\mathcal H_S$, with two dissipative evolutions described by the local Lindbladians $\mathcal L_1$ and $\mathcal L_2$.
As discussed in point~\ref{it:LR bounds} in Section~\ref{subsec:formal def}, they are local in the sense that they can be written as sums of local terms, $\mathcal L_a = \sum_X \mathcal L_X^{(a)}$ with $\supp[ \mathcal L_X^{(a)}] = X$.
We ask the $X$'s to be the subsets of the lattice with size at most $k$, constant in $N$.
In this case the norm of the local terms will be independent of the size of the support, and at most of order one, $\norm*{\mathcal L_X^{(a)}} \lesssim O(1)$.
However, we can also extend the definition of locality by letting $X$ to be of any size, and constraining the norm of $\mathcal L_X^{(a)}$ to be smaller than some decaying function of $|X|$, following Ref.~\cite{Bachmann2012}.
In both cases $\norm*{\mathcal L}\lesssim O(N)$, with $N$ the size of the system.

As a second step, we add identical ancillary timers to every point of the system, living in the Hilbert space $\mathcal H_T = \bigotimes_{i=1}^N\mathcal H_{T(i)}$.
Following~\cite{Kastoryano2013}, the timer leaving in $\mathcal H_{T(i)}$ is composed of $T+1$ qubits whose evolution is described by the following Lindbladian operator
\begin{equation}
\label{eq:timer evolution}
\mathcal L_{T(i)}(\rho) = \sum_{j=0}^T L_j \rho L_j^\dag - \frac{1}{2} \{L_j^\dag L_j,\rho\}, \quad 
L_j = \sqrt\gamma \ketbrasub{1}{1}{j}\otimes \ketbrasub{1}{0}{j+1}.
\end{equation}
Here we drop the dependence on $i$ of each Lindblad operators of the timer, since it shouldn't generate any confusion.
Each timer is initialized such that all qubits are in $\ket{0}$, but the first one $\ket{\phi_0}=\ket{1}_0\otimes\left(\bigotimes_{j=1}^T \ket{0}_j\right)$.
Each timer evolves independently according to~\eqref{eq:timer evolution}, by flipping the qubits one by one.
The only states accessible by this evolutions are the ones of the form $\ket{\phi_k}=\left(\bigotimes_{j=0}^k\ket{1}_j\right)\otimes\left(\bigotimes_{j=k+1}^T\ket{0}_j\right)$.
We also define the projectors $\phi_k = \ketbra{\phi_k}{\phi_k}$.
If $T$ is large enough, the last qubit will be flipped with high probability at a time $\tau=T/\gamma$.
This last qubit is the switch and it will control the Lindbladian acting on the original system.
In \cite{Kastoryano2013} it is shown that the probability for the switch to be in the state $\ket{1}_T$ before $\tau$ is given by
\begin{equation}
p_1^T(t<\tau) = e^{-\frac{t}{\tau}T}\sum_{j=T}^\infty \left(\frac{t}{\tau}\right)^j\frac{T^j}{j!} = 
e^{-\frac{t}{\tau}T}\left(\frac{t}{\tau}\right)^T\frac{T^T}{T!}\sum_{j=0}^\infty \left(\frac{t}{\tau}\right)^j \frac{T^j\,T!}{(T+j)!}.
\end{equation}
The last factor in the argument of the sum can be written as $\prod_{k=1}^{j}\left(1+k/T\right)^{-1}$ and therefore it is always smaller than one.
Using Stirling's approximation on the factor that multiply the sum, it is easy to see that if $t=\tau-\eps/2$, and for small $\eps$,
\begin{equation}
\label{eq:exp error 1}
p_1^T(t) \lesssim \frac{e^{-T\left(\frac{t}{\tau}-\log\frac{t}{\tau}-1\right)}}{\sqrt{2\pi T}}\frac{1}{1-t/\tau} 
\lesssim \experr .
\end{equation}
The probability of the switch to be in the state $\ket{1}_T$ in $t=\tau-\eps/2$ goes to zero exponentially for $T\to\infty$ if $\eps\to 0$, in a way such that the combination $\sqrt T\eps / (2\tau) \rightarrow\infty$.
This is satisfied if for example we choose $\eps$ to scale with $T$ as $\eps \sim T^{-\alpha}$, with $0<\alpha<1/2$.
Analogously we can see that the probability of the switch being in the state $\ket{0}_T$ at time $t=\tau+\eps/2$ is vanishing exponentially with the same scaling of $\eps$ with the size $T$ of the timer.
Indeed,
\begin{equation}
p_0^T(t>\tau) = e^{-\frac{t}{\tau}T}\sum_{j=0}^{T-1} \left(\frac{t}{\tau}\right)^j\frac{T^j}{j!} = 
e^{-\frac{t}{\tau}T}\left(\frac{t}{\tau}\right)^{T-1}\frac{T^T}{T!}\sum_{j=0}^{T-1} \left(\frac{\tau}{t}\right)^j \frac{T^{-j}(T-1)!}{(T-j-1)!}.
\end{equation}
This time the last factor in the argument of the sum can be written as $\prod_{k=1}^{j}\left(1-k/T\right)<1$.
Using again Stirling's approximation, we find that if $t=\tau+\eps/2$,
\begin{equation}
\label{eq:exp error 2}
p_0^T(t) \lesssim \frac{e^{-T\left(\frac{t}{\tau}-\log\frac{t}{\tau}-1\right)}}{\sqrt{2\pi T}}\frac{1}{t/\tau-1}
\lesssim \experr .
\end{equation}
Let us make here a notational comment:
Here we used the symbol '$\lesssim$' to indicate that the upper bound is true up to multiplicative constant (in the variables relevant to the problem) or to subleading contributions in the large $N$ limit.
Indeed, we will see later that the size of the ancilla space must increase in the thermodynamic limit, therefore the large $T$ limit is not independent from the large $N$ limit.
In the following, we will also use the symbol '$\simeq$'  to indicate equality up to  additive error terms that vanish fast in the thermodynamic limit, as for example the ones discussed in  eqs.~\ref{eq:exp error 1} and \ref{eq:exp error 2}.
Note that $\simeq$ is {\it not} equivalent then to $\lesssim$ and $\gtrsim$.

As already mentioned, the original system is coupled with the switches of the timer, in a way that they act as a control on the evolution.
For every local term in the Lindbladians $\mathcal L^{(a)}_{X}$ choose a site $i\in X$ and consider the new Lindbladian for the system coupled to the timers $\mathcal L_S = \sum_X \mathcal L_X$, where each term $\mathcal L_X$ is defined through the Lindblad operators $L_{X,j} = \ketbrasub{0}{0}{T}\otimes L_{X,j}^{(1)} + \ketbrasub{1}{1}{T}\otimes L_{X,j}^{(2)}$ and the Hamiltonian $H_X = \ketbrasub{0}{0}{T}\otimes H_X^{(1)} + \ketbrasub{1}{1}{T}\otimes H_X^{(2)}$.
Here $L_{X,j}^{(a)}$ and $H_X^{(a)}$ for $a=1,2$ are the Lindblad operators and the Hamiltonians of the original Lindbladians $\mathcal L_1$ and $\mathcal L_2$, and the switch considered is the one of the timer attached at site $i$.
Until all the switches are in the state $\ket{0}_T$ the system evolves with $\mathcal L_1$, after they all flip to $\ket{1}_T$ the evolution is governed by $\mathcal L_2$.
The full evolution of the system and the ancillas is governed by $\mathcal L = \mathcal L_S + \mathcal L_T$.

We will now prove that the effective evolution of the system described by $\mathcal L$ is equivalent of evolving with $\mathcal L_1$ for $t<\tau$ and with $\mathcal L_2$ for $t>\tau$.
The strategy will consist of three steps.
We will show that:
\begin{enumerate*}[label=(\textit{\roman*})]
\item \label{it:before tau}
The evolution of a state $\rho$ under $\mathcal L$ is almost equal to the one under $\mathcal L_1$ for $t<\tau^-=\tau-\eps/2$ if $\eps$ scales with $T$ as described before;
\item \label{it:window}
After tracing out the ancillas, the system evolved for a time $\eps$ is $\eps$-close to the initial state;
\item \label{it:after tau}
At $t>\tau^+=\tau+\eps/2$ with high probability the timers are all in the fixed point of $\mathcal L_T$ where all the switches are in $\ket{1}_T$, and the evolution of the system is obtained up to vanishing errors by evolving with $\mathcal L_2$.
\end{enumerate*}

\paragraph{First time interval, $t<\tau^-$.}
At time $t=0$ all clocks are initialized in the state $\phi_0 = \bigotimes_{i=1}^N \phi_0^{(i)}$, while the system is in $\rho_0$.
The key to prove step~\ref{it:before tau} is to show that the state at any time can be written as follows
\begin{equation}
\label{eq:before tau}
e^{t\mathcal L}(\phi_0\otimes\rho_0) = \sum_{k_1,\dots,k_N = 0}^{T-1} p_{\{k_i\}}(t)\,\phi_{\{k_i\}} \otimes e^{t\mathcal L_1}(\rho_0) 
+ \sideset{}{'}\sum_{\{k_i\}} p_{\{k_i\}}(t)\,\phi_{\{k_i\}}\otimes \rho_{\{k_i\}}'.
\end{equation}
Here $p_{\{k_i\}}$ is the probability of the timers being in the state $\phi_{\{k_i\}}=\bigotimes_i \phi_{k_i}$, and since the timers are all identical and independent $p_{\{k_i\}}=\prod_i p_{k_i}$, with $p_{k}$ the probability of a single timer to be in the state $\phi_{k}$.
The primed sum is a sum over all $\{k_i\}_i$ with at least one $k_j = T$, and $\rho_{\{k_i\}}'$ is some state that depends on the $k_i$'s.
Define the projector $P_{T} = \sum_{k_1,\dots,k_N = 0}^{T-1}\phi_{\{k_i\}}$, and the corresponding superoperator $\mathcal P_{T}(A) = P_{T}\,A\, P_{T}$.
By direct inspection one can see that $\mathcal P_{T}\circ\mathcal L\circ\mathcal P_{T} (\rho) = \mathcal P_{T}\circ\mathcal L(\rho)$.
Moreover, since $\mathcal P_{T}\circ\mathcal L_S\circ\mathcal P_{T} = \mathcal P_{T}\otimes \mathcal L_1$, we have 
$\mathcal P_{T}\circ\mathcal L\circ\mathcal P_{T} (\rho) = \mathcal P_{T}\circ(\mathcal L_T + \mathcal L_1)(\rho)$.
Applying this result to $\mathcal P_{T} e^{t\mathcal L}(\phi_0\otimes\rho_0)$, it is immediate to see that it is equal to $\mathcal P_{T} e^{t\mathcal L_T}(\phi_0)\otimes e^{t\mathcal L_1}(\rho_0)$, and from this conclude~\eqref{eq:before tau}.
Finally, if we specialize eq.~\eqref{eq:before tau} to $\tau^-=\tau-\eps/2$ and we trace out the ancillas
\begin{equation}
\label{eq:before tau trace}
\Tr_T\left[ e^{\tau^-\mathcal L}(\phi_0\otimes\rho_0) \right] =
\left[\sum_{k_1,\dots,k_N=0}^{T-1} p_{\{k_i\}}(\tau^-)\right] e^{\tau^-\mathcal L_1}(\rho_0)
+ \sideset{}{'}\sum_{\{k_i\}} p_{\{k_i\}}(\tau^-) \rho_{\{k_i\}}'.
\end{equation}
The factor in parenthesis is the probability of all switches being in the state $\ket{0}_T$ at $\tau^-$
\begin{equation}
\label{eq:p0 Ntimers}
p_0(\tau^-) = \prod_{i=1}^N p_0^{T}(\tau^-) 
\simeq \left[1-\experr\right]^N
\simeq 1- \experr[N].
\end{equation}
On the other end, the second term of eq.~\eqref{eq:before tau trace} has norm one of order $\sim 1-p_0(\tau^-)$.
This probabilities go exponentially to one and zero if $\eps$ scales with $T$ as discussed.
We also need to scale $T$ with the system size, for example as a power law $T\sim N^\beta$, with an exponent $\beta$ to be determined later.
Therefore in the thermodynamic limit
\begin{equation}
\norm{\Tr_T\left[ e^{\tau^-\mathcal L}(\phi_0\otimes\rho_0) \right] - e^{\tau^-\mathcal L_1}(\phi_0)}_1 \xrightarrow[N\to\infty]{} 0 ,
\end{equation}
and the limit is exponentially fast.

Let us now make a couple of comments about the engineering of this kind of timer.
The timer needs to flip the switch around a specific given time $\tau$, and this event must happen with high probability in a very small time window.
The error and the width of the time window must vanish in the thermodynamic limit.
As noticed before, the time scale $\tau$ is given by $T/\gamma$, and we just saw that $T$ must scale (polynomially) with the system size $N$.
Then, apart from some $\poly\log N$ terms, the parameter $\gamma$ must scale as $\poly N$ too.
It is a bit unsettling to have the need for the size of the timer and the coupling constant governing its evolution, to be tuned according to the original system size.
However, it is also clear that to keep locality we need $\sim O(N)$ timers and therefore the error depends on $N$, as seen in eq.~\eqref{eq:p0 Ntimers}.
Moreover, the timer of the kind used in~\cite{Kastoryano2013} triggers an event by sequentially flipping all the ancillas.
We expect the time needed for this process to be proportional to the size of the timer $T$.
In order for the event to be triggered at a finite time $\tau$, the flipping process must happen very fast and therefore the coupling constant $\gamma$ must be very large.
It would be interesting to devise other timers, with different characteristics.
One interesting possibility would be to look for a timer with a larger size (but still $\poly N$), but with coupling constants of the same order of the ones in the original system.

It is also worth commenting on the size of the ancillas' space.
Each timer is composed of $T\sim N^\beta$ qubits and its evolution is governed by a Lindbladian with 2-qubit interactions, apart from the last one, which interacts with the sites in some subset of the lattice where the original system lives.
Therefore the dimension of the ancillas' Hilbert space is exponential in $N$, i.e.\ $\dim\mathcal H_{T(i)}\sim 2^{N^\beta}$.
However, only the $T$ states $\phi_k$ are actually accessible by the evolution, so the effective total dimension is lower.
For example, if we block all the ancillas together we could get only one qudit with $d=T$.
We can visualize this situation as a single system with $T$ equally spaced excited states, initialized in the highest energy level and with decay rate $\gamma$ from any state to the directly lower one (and considerably smaller decay rates for decay processes which involve jumps of more than one level).
Another point of view could be to block the original qubits in a way to end up with $\sim\log_2 T$ qubits whose states are all accessible by the dynamics.
In this case we would loose the two-locality of the interactions between the timer ancillas, however the interaction terms would involve at most $\log T\sim \log N$ of the blocked qubits.
This is still acceptable for the reasons mentioned in observation~\ref{it:LR bounds} in Section~\ref{subsec:formal def}.
Notice that this would make the Lieb-Robinson bounds in the timer trivial, but it wouldn't have any effects on the original system, since each timer is connected to a single site.
We also anticipate that in the application of the timer in Section~\ref{sec:pure case}, the interactions in the system are already supported on regions of size $\poly\log N$.
Therefore any $\log$-locality in the timer seems very natural.

\paragraph{Second time interval, $\tau^-<t<\tau^+$.}
In the second step in the proof of transitivity we focus on the time window around $\tau$.
We ultimately want to show that 
\begin{equation}
\label{eq:window}
\norm{ \Tr_T\left[e^{\tau^+\mathcal L}(\phi_0\otimes\rho_0) - e^{\tau^-\mathcal L}(\phi_0\otimes\rho_0)\right] }_1
\lesssim \eps \norm*{\mathcal L_S}_{1\rightarrow 1} 
\lesssim \eps N \max_{X,a\in{1,2}} \norm*{\mathcal L_X^{(a)}}_{1\rightarrow 1}.
\end{equation}
This difference goes to zero if $\eps N \rightarrow 0$, and this happens if for example $\beta>1/\alpha$.
Since $0<\alpha<1/2$, notice than in any case we can never obtain a vanishing error if $\beta\leq 2$.
To show~\eqref{eq:window} we use a general argument which formalizes an intuitive fact on the dynamics of a system weakly coupled with a second system whose time scales are much faster (in the case at hand the latter is the timer).
Weakly coupled here means that the coupling constants between the two systems are at most of the same order of magnitude as the ones that govern the dynamics of the original system (for concreteness, let us consider them of order one).
If the fast system is traced out, the resulting dynamics must loose track of the stronger coupling constants and the time scales must then be of order one.
More precisely, consider a Hilbert spaces$\mathcal H = \mathcal H_T \otimes \mathcal H_S$ and a Lindbladian $\mathcal L = \mathcal L_T + \mathcal L_S$, where $\mathcal L_S$ acts on the full Hilbert space $\mathcal H$ and has a norm $\norm*{\mathcal L_S}_{1\rightarrow 1}\sim O(N)$, where $N$ is the size of the system living in $\mathcal H_S$.
On the contrary $\mathcal L_T$ acts only on $\mathcal H_T$ and its norm is much larger than $\norm*{\mathcal L_S}_{1\rightarrow 1}$, namely $\norm*{\mathcal L_T}_{1\rightarrow 1}\sim \omega(N)$.
This large norm can be due to the fact that $\mathcal H_T$ is a huge bath compared to the size of the system, or that its internal dynamics is very fast.
Consider the evolution of the system $\rho_S(t) = \Tr_T\left[e^{t\mathcal L}(\rho)\right]$.
If we evolve for a time $\eps$, which is small compared to the time scales of $\mathcal L_S$, the evolution should be $\eps$-close to the identity, even if the time scales of the bath $T$ are much shorter than $\eps$ and the final state of the bath is potentially very different from the one at the beginning of this time interval.
To estimate $\norm{\rho_S(\eps) - \rho_S(0)}_1$ we need to estimate $\norm{\partial_t\rho(t)}_1$, indeed
\begin{equation}
\label{eq:general argument}
\begin{split}
\norm{\rho_S(\eps) - \rho_S(0)}_1 
&= \sup_{\norm{\Theta}=1} \Tr\Big[\Theta\,\big(\rho_S(\eps)-\rho_S(0)\big)\Big] \\
&= \eps \,\sup_{\norm{\Theta}=1} \Tr\left[\Theta\,\partial_t\rho_S(t)\Big\vert_{\substack{t=\bar t\\0<\bar t<\eps}}\right] 
 = \eps \,\norm{\partial_t\rho_S(\bar t)}_1.
\end{split}
\end{equation}
Now, $\partial_t\rho_S(t) = \Tr_T\left[\left(\mathcal L_T + \mathcal L_S\right) e^{t\mathcal L}(\rho)\right]$.
The trace of any Lindbladian applied to any state is traceless, however in this case $\mathcal L_T$ acts non trivially only on $\mathcal H_T$, and as the identity on $\mathcal H_S$.
Also, the trace is just a partial trace on the bath.
To check explicitly that the first term of the sum vanishes let us show that $\Tr_T\left[\mathcal L_T(\rho)\right]$ vanishes as an operator in $\mathcal H_S$.
For any bounded operator $\Theta\in\mathcal B(\mathcal H_S)$,
\begin{equation}
\Tr_S\left[\Theta\,\Tr_T\left[\mathcal L_T(\rho)\right]\right] = \Tr\left[(\mathbb{I}_T\otimes\Theta)(\mathcal L_T\otimes\mathbb{I}_S)(\rho)\right] = 
\Tr\left[\rho(\mathcal L_T^*\otimes\mathbb{I}_S)(\mathbb{I}_T\otimes\Theta)\right]=0,
\end{equation}
since $\mathcal L_T^*(\mathbb{I}_T)=0$.
Finally a bound on eq.~\eqref{eq:general argument} is obtained,
\begin{equation}
\norm{\partial_t\rho_S(t)}_1 = \norm{\Tr_T\left[\mathcal L_S\, e^{t\mathcal L}(\rho)\right]}_1 \leq \norm*{\mathcal L_S}_{1\rightarrow 1}\norm*{e^{t\mathcal L}(\rho)}_1
= \norm*{\mathcal L_S}_{1\rightarrow 1}.
\end{equation}
If we now rewrite the lhs of eq.~\eqref{eq:window} as $\norm{\Tr_T\left[e^{\eps \mathcal L}(\rho(\tau^-))\right] - \Tr_T\left[\rho(\tau^-)\right]}_1$, we can immediately apply the last result and obtain the rhs.

\paragraph{Third time interval $\tau>\tau^+$.}
The final step consists in studying what happens after $\tau^+=\tau+\eps/2$.
With exponentially high probability the timers are in the state where all switches have been flipped to one, namely $\phi_T=\bigotimes_{i=1}^N\phi_T^{(i)}=\bigotimes_{i=1}^N\ketbra{1}^{\otimes T+1}$, which is a fixed point of the evolution described by $\mathcal L_T$.
Therefore the only effect of $\mathcal L$ is to act on $\mathcal H_S$ as $\mathcal L_2$.
More precisely, the global state for $t\geq \tau^+$ can be written as follows:
\begin{equation}
\label{eq:transitivity after tau}
e^{t\mathcal L}(\phi_0\otimes\rho_0) = 
\sum_{\{k_i\}_i\neq \{T\}_i} p_{\{k_i\}}(t)\, \phi_{\{k_i\}} \otimes \rho_{\{k_i\}}'' + \prod_{i=1}^N p_1^T(t)\, \phi_T\otimes \rho'(t).
\end{equation}
The product in the second term is  exponentially close to one, 
$p_T(\tau^+)\simeq 1-\experr[N]$,
therefore the state is exponentially close to the second factor in the sum
\begin{equation}
e^{t\mathcal L}(\phi_0\otimes\rho_0) = \phi_T \otimes \rho'(t)
+ \err{\experr[N]}.
\end{equation}
From the result of step~\ref{it:window} we know that at $t=\tau^+$ the state in $S$ is close to the state at $t=\tau^-$, therefore
\begin{equation}
\rho'(\tau^+) = \Tr_T\left[e^{\tau^-\mathcal L}(\phi_0\otimes\rho_0)\right]
+ \err{\eps N,\experr[N]}.
\end{equation}
And from the results of step~\ref{it:before tau} we can conclude that $\rho'(\tau^+) \simeq e^{\tau^-\mathcal L_1}(\rho_0) + \err{\eps N}$, where we kept only the largest error term (in the thermodynamic limit).
This implies that the state at times $t>\tau^+$ is given by
\begin{equation}
\label{eq:error}
\begin{split}
e^{t \mathcal L}(\phi_0&\otimes\rho_0)
= e^{(t-\tau^+) \mathcal L}e^{\tau^+ \mathcal L}(\phi_0\otimes\rho_0) 
\simeq e^{(t-\tau^+) \mathcal L}\left( \phi_T\otimes e^{\tau^-\mathcal L_1}(\rho_0) \right) + \err{\eps N} \\
&\simeq \phi_T \otimes e^{(t-\tau^+)\mathcal L_2}e^{\tau^-\mathcal L_1}(\rho_0) + \err{\eps N} 
\simeq \phi_T \otimes e^{(t-\tau)\mathcal L_2}e^{\tau\mathcal L_1}(\rho_0) + \err{\eps N}.
\end{split}
\end{equation}
In the last equation, we simply substituted $\tau^\pm$ with $\tau$ since this will only introduce an error term of order $\err{\eps N \norm{\mathcal L_S}}$.
We also used the fact that the Lindbladian evolution is norm-1 contractive, and all error terms must be understood as operators with 1-norm of that order.

\paragraph{Transitivity.}
To finish the discussion about transitivity, let us show that the Lindbladian $\mathcal L$ does indeed drive $\rho_0$ to $\rho_2$ fast, namely $\rho_0\fastto[\mathcal L]\rho_2$.
By hypothesis we have that $\rho_0\fastto[\mathcal L_1]\rho_1$ and $\rho_1\fastto[\mathcal L_2]\rho_2$, which more explicitly means that for times $t_1, t_2 \gtrsim \poly\log N$,
\begin{equation}
\label{eq:transitivity hyp}
\norm{e^{t_1\mathcal L_1}(\rho_0)-\rho_1}_1 \leq \eps_{N}^{(1)}, \qquad
\norm{e^{t_2\mathcal L_2}(\rho_1)-\rho_2}_1 \leq \eps_{N}^{(2)}.
\end{equation}
Let us now set the timer parameters such that $\tau=t_1$ and define $t=t_1+t_2$.
Then,
\begin{equation}
\label{eq:transitivity argument}
\begin{split}
&\norm{\Tr_T\left[e^{t\mathcal L}(\phi_0\otimes\rho_0)\right]-\rho_2}_1 
\simeq \norm{e^{(t-\tau)\mathcal L_2}e^{\tau \mathcal L_1}(\rho_0)-\rho_2}_1 + \err{\eps N} \\
&\lesssim \norm{e^{(t-\tau)\mathcal L_2}e^{\tau \mathcal L_1}(\rho_0) - e^{(t-\tau)\mathcal L_2}(\rho_1)}_1 + \norm{e^{(t-\tau)\mathcal L_2}(\rho_1)-\rho_2}_1 + \err{\eps N} \\
&\lesssim \norm{e^{\tau \mathcal L_1}(\rho_0) - \rho_1}_1 + \norm{e^{(t-\tau)\mathcal L_2}(\rho_1)-\rho_2}_1 + \err{\eps N}
\lesssim \eps_{N}^{(1)} + \eps_{N}^{(2)} + \err{\eps N},
\end{split}
\end{equation}
where we used the hypothesis eq.~\eqref{eq:transitivity hyp}, triangle inequality, and the fact that the Lindbladian evolution is norm-1 contractive.
We kept only the largest term (in the thermodynamic limit) of the contributions to the error in eq.~\eqref{eq:error}.
The right hand side of eq.~\eqref{eq:transitivity argument} goes to zero as $N\to\infty$ and this proves transitivity.

\section{The pure state case, a path of gapped Hamiltonians implies a fast Lindbladian evolution}
\label{sec:pure case}

In the rest of this work we consider the case of pure states and explore which kind of classification of phases is implied by the definition through fast Lindbladian evolution.
In particular we would like to understand if this definition is equivalent to the usual one with a path of gapped Hamiltonians.
We won't be able to give a complete answer, however in this section we show that if for two pure states such a gapped path exists, then we can build a $\log$-local Lindbladian (meaning with terms whose support has size $\poly\log N$) which drives one state to the other in time $\poly\log N$.

First of all, let us recall informally the definition of gapped phases valid for pure states through the path of gapped Hamiltonians. The concrete hypothesis required on the path of Hamiltonians in this section will be made explicit at the beginning of Section \ref{subsec:finite depth circuit}. 
\begin{definition}[Hamiltonian phases]
\label{def:H definition}
Two pure states $\ket{\psi_0}$ and $\ket{\psi_1}$ are in the same phase if there exist two local gapped Hamiltonian $H_0$ and $H_1$ such that $\ket{\psi_0}$ is the ground state of $H_0$ and $\ket{\psi_1}$ is the ground state of $H_1$, and a \emph{sufficiently regular} path $H(s)$ of local Hamiltonians such that $H(0)=H_0$, $H(1)=H_1$ and $H(s)$ is gapped uniformly in $s$, namely the gap $\lambda(s)$ between the ground state space and the first excited level of $H(s)$, is uniformly bounded by a constant in the thermodynamic limit, $\lambda(s)\geq \lambda > 0$.
\end{definition}
Notice that given such a path $H(s)$, for any different choice of gapped Hamiltonians $H_0'$ and $H_1'$ whose ground states are $\ket{\psi_0}$ and $\ket{\psi_1}$, a gapped path $H'(s)$ such that $H'(0)=H_0$ and $H'(1)=H_1$ exists.
Indeed it is possible to build a gapped path connecting $H_0$ and $H_0'$ by simple linear interpolation $H_0(s)=(1-s)H_0 + s H_0'$.
When taking the point of view of states, it is important to make a remark on the choice of the gapped Hamiltonian $H_0$ whose ground state is $\ket{\psi_0}$.
Indeed, we must choose $H_0$ such that it is the one with the minimal size of the ground state space.
For example, for states in some non-trivial phase we can only choose $H_0$ with a degenerate ground state space.
Such ground state space could contain a product state, however for this product state it is possible to find a Hamiltonian for which it is the unique ground state.
The latter should be chosen as the starting point of any gapped path from the product state.

Notice also that Definition~\ref{def:H definition} can be modified allowing for some ancillas to be added to the Hilbert space.
These ancillas must act as catalyser and have similar characteristics to the locality preserving ancillas defined in observation~\ref{it:ancillas} in Section~\ref{sec:definition}, but subject to a Hamiltonian evolution.
We do not expect the classification of phases for pure states to change with this addition to the definition, however the latter is necessary when comparing systems with different local dimensions.
Indeed, we will need to consider ancillas in the definition of the gapped path in Section~\ref{subsec:1D SPT phase}, when studying Symmetry Protected Topological (SPT) phases in one dimension.
In that setting, the ancillas are necessary to connect states that transform according to different representations of the same symmetry group.

We can now state the main result of this section.
\begin{main}
Given two states $\ket{\psi_0}$ and $\ket{\psi_1}$ that are in the same phase according to the Hamiltonian Definition~\ref{def:H definition}, they also are in the same phase according to the Lindbladian Definition~\ref{def:L definition}.
In our construction, the Lindbladian $\mathcal L$ describing the dissipative evolution from $\ketbra{\psi_0}$ to $\ketbra{\psi_1}$ is $\log$-local, meaning it is a sum of terms each acting on a subset of the lattice of size $\poly\log N$.
\end{main}

The strategy we follow consists of mainly two steps.
In the first step we consider the quasi-adiabatic Hamiltonian which describes a time-dependent evolution which unitary connects two pure states in the same phase.
The existence of such an evolution is a direct consequence of the path of gapped Hamiltonians $H(s)$ that connects the two pure states.
We show that such a unitary evolution can be approximated with a finite depth quantum circuit with gates that act on blocks of spins of size $\poly\log N$.
The second step consists in building a Lindbladian that effectively implements this quantum circuit with the help of the ancillary timers already introduced in Section~\ref{subsec:transitivity}.

\subsection{Adiabatic evolution as a $\poly\log$-local circuit of constant depth}
\label{subsec:finite depth circuit}

As mentioned in the Introduction, precise results about approximating the Hamiltonian evolution with a finite depth quantum circuit are available for the very general cases of a strictly local Hamiltonian (or one that is a sum of exponentially decaying terms).
In~Ref.\cite{Osborne2005} the author proves that the evolution of a local one dimensional Hamiltonian can be approximated by a finite depth circuit with gates acting on regions of size logarithmic in the system size.
In~Ref.\cite{Haah2018} the result is extended to higher dimensions, and the resulting circuit is further refined using known Hamiltonian simulation algorithms to obtain strictly local gates.
As a trade-off, the depth of this new circuit is $\poly$-logarithmic in the system size.
See also Ref.~\cite{Tran2018a} for the analysis of Hamiltonians with polynomially decaying terms.
Unfortunately the quasi-adiabatic Hamiltonian is only quasi-local~\cite{Bachmann2012}, hence these results cannot be applied directly.
The case of the quasi-adiabatic Hamiltonian is considered in Refs.~\cite{Osborne2007,Huang2015}, where the authors look for approximations of the local density matrices.
Since we are interested in the full state, then again we cannot directly apply their results here.
However, by following the same line of thoughts of these works, we can obtain the approximation and the precise error bounds that we will need later on.

For the sake of completeness, let us show how it is possible to extend the result of Ref.~\cite{Osborne2005} to the case of the quasi adiabatic Hamiltonian $K(t)$ corresponding to the path $H(s)$.
The main difference with what proven in Ref.~\cite{Osborne2005} is that $K(t)$ is quasi-local.
We need to use the results of Ref.~\cite{Bachmann2012} about precise statements on the Lieb-Robinson bounds that $K(t)$ satisfies.
We will also use techniques from Ref.~\cite{Osborne2007}.
We will approximate the quasi-adiabatic evolution with a circuit with gates acting on regions of size $\poly\log N$.
Let us notice that using some arguments in Ref.~\cite{Haah2018} the result could be extended, at least in certain cases, to a circuit made of strictly local gates and with $\poly\log$ depth.
However, since we do not require strict locality, we will not dwell into the details of when this refinement can actually be performed.

\begin{step}
For any two states $\ket{\psi_0}$ and $\ket{\psi_1}$ that are in the same Hamiltonian phase (according to Definition~\ref{def:H definition}), there exists a unitary circuit $C$ of finite depth (independent of the system size) and with gates acting on regions of size $\poly\log N$, such that 
\begin{equation}
\norm{C\ketbra{\psi_0}C^\dag-\ketbra{\psi_1}}\leq \delta_N,
\end{equation}
with $\delta_N$ vanishing for $N\to\infty$ faster than any power.
\end{step}

Consider the two pure states $\rho_0=\ketbra{\psi_0}$ and $\rho_1=\ketbra{\psi_1}$ and two local Hamiltonians $H_0$ and $H_1$ such that $\Tr\left[H_0\rho_0\right]=0=\Tr\left[H_1\rho_1\right]$.
If $\rho_0$ ad $\rho_1$ are in the same phase, then there exists a path of gapped local Hamiltonians $H(s)$, $s\in [0,1]$ such that $H(0)=H_0$ and $H(1)=H_1$.

The Hamiltonian path $H(s)$ may also be chosen to be quasi-local.
Since the quasi-adiabatic Hamiltonian is in any case quasi-local, we might as well consider a quasi-local Hamiltonian path $H(s)$.
This means that it is a sum of terms with norms that depend on the size of their support.
For concreteness, consider a lattice $\Lambda$ with dimensionality $d$, and all the balls $B_{j,\alpha}=\{i\in\Lambda\mid d_{i,j}<\alpha \}$, centred in $j$ and with radius $\alpha$.
We will assume that the structure of $\Lambda$ is such that $\abs{B_{j,\alpha}}\leq \kappa \alpha^d$ for some constant $\kappa$, for example we can consider a square lattice.
Then, following Ref.~\cite{Bachmann2012}, the Hamiltonian can be written as
\begin{equation}
H(s)=\sum_{j\in\Lambda}\sum_{\alpha\geq 1}h_{j,\alpha}(s),\quad\text{with}\quad\supp\left[h_{j,\alpha}(s)\right]=B_{j,\alpha}.
\end{equation}
Quasi-locality means that there exists a function $F$ such that
\begin{equation}
\label{eq:quasilocality}
\norm{h}_F = \sup_{x,y\in\Lambda}\frac{1}{F(d_{x,y})}\sum_{\substack{j,\alpha:\\x,y\in B_{j,\alpha}}}\norm{h_{j,\alpha}} < \infty,
\end{equation}
where $d_{x,y}$ is the distance between $x$ and $y$ in the lattice $\Lambda$, and we dropped the dependence on the parameter $s$, since all bounds are going to be uniform in $s$.
To apply the results of Ref.~\cite{Bachmann2012} we need $F$ to be of the form $F_a(x)=e^{-a x}F(x)$ and in this case we write $\norm{h}_a$.
The function $F(x)$ must satisfy an integrability property and a convolution property,
\begin{gather}
\label{eq:F integrability}
\norm{F} = \sup_{y\in\Lambda}\sum_{x\in\Lambda}F(d_{x,y}) < \infty, \\
\label{eq:F convolution}
\sum_{z\in\Lambda} F(d_{x,z})F(d_{z,y}) \leq C_F F(d_{x,y}).
\end{gather}
For example a possible choice is $F(x)=(1+x)^{-(d+1)}$.
A direct consequence of eq.~\eqref{eq:quasilocality} is that each term in the Hamiltonian is bounded by $\norm{h_{j,\alpha}}\leq \norm{h}_a F_a(2\alpha)$ and that $\norm{H}\leq N \norm{h}_a F(0)$, with $N=|\Lambda|$ the size of the system.
The path of Hamiltonian is also needed to be smooth with bounded derivative, namely $\norm{\partial h}_b<\infty$, with $\partial h_{j,\alpha}=|B_{j,\alpha}|h_{j,\alpha}'(s)$. We will also assume that there is uniform exponential Lieb-Robinson bound throughout the path (assumption 4.4 in \cite{Bachmann2012}). Namely
\begin{equation}
\label{eq:Assumption4.4}
\left\| \left[ \tau_t^{H_\Lambda(s)}(A),B\right] \right\|\le K_a\|A\|\|B\|e^{av_a|t|}\sum_{x,y} F_a(d_{x,y}) 
\end{equation}
where the sum is on $x$ (resp.\ $y$) belonging to the support of $A$ (resp.\ $B$) and the constants $v_a, K_a$ are independent of $s$ and $\Lambda$.

Given such path of gapped Hamiltonians, it is possible to define the quasi-adiabatic Hamiltonian $K(s)$,
\begin{equation}
\label{eq:quasiadiabatic K}
K(s) = \int_{-\infty}^{\infty} \dd{t} W_\lambda(t)\,\tau_t^{H(s)}\left(H'(s)\right),
\end{equation}
where $\tau_t^{H}$ is the unitary evolution for a time $t$ with the Hamiltonian $H$, namely $\tau_t^H(A)=e^{\iu H}A\,e^{\iu H}$, and $W_\lambda(t)$ is a filter function as defined in Ref.~\cite{Bachmann2012} and depends on the uniform gap $\lambda$.
The quasi-adiabatic Hamiltonian turns out to be quasi-local and gapped at any time $s$.
We call $\alpha_t$ the unitary evolution defined by $K(t)$, $\alpha_t(\rho_0) = e^{\iu \int_0^t \dd{s} K(s)}\,\rho_0\, e^{-\iu \int_0^t \dd{s} K(s)}$.
It can be shown that $\alpha_1(\rho_0) = \rho_1$.

\subsubsection{Quasi-locality of the quasi-adiabatic Hamiltonian}
\label{subsubsec:K quasi-local}

The quasi-adiabatic Hamiltonian is shown to be quasi-local in very general terms in Ref.~\cite{Bachmann2012}.
It is also shown in Ref.~\cite{Osborne2007} in a setting fitting more with ours, but the results are not strong enough for our purposes.
Let us then follow the derivation of Ref.~\cite{Osborne2007}, while improving the bounds using some results of Ref.~\cite{Bachmann2012} on the filter functions.

\begin{substep}
The quasi adiabatic Hamiltonian can be written as a sum of terms
\begin{equation}
K(s) = \sum_{j\in\Lambda} \sum_{\alpha=1}^{L} k_{j,\alpha}(s),
\end{equation}
such that $k_{j,\alpha}(s)$ is supported on $B_{j,\alpha}$, namely the ball centred in $j$ and of radius $\alpha$.
Moreover, the following bound on the norm of $k_{j,\alpha}$ holds,
\begin{equation}
\norm{k_{j,\alpha}(s)} \lesssim \alpha\,I_\lambda\left(b\,\alpha\right),
\end{equation}
where $b$ is a constant, and $I_\lambda(t)\lesssim (\lambda t)^{10} u_{2/7}(\lambda t)$ with $u_\mu(t)=e^{-\mu\frac{t}{\log^2(t)}}$.
\end{substep}
As already commented, the symbol '$\lesssim$' means that the bound is valid up to irrelevant constants that we won't keep track of and up to subleading contributions in the thermodynamic limit.

The quasi-adiabatic Hamiltonian can be rewritten as follows,
\begin{equation}
\label{eq:quasiadiabatic Delta}
K(s) = \sum_{j,\alpha}\int_{-\infty}^{\infty} \dd{t} W_\lambda(t)\,\tau_t^{H(s)}\left(h'_{j,\alpha}(s)\right) = \sum_{j,\alpha}\sum_{n\geq 0} \Delta^n\left(h'_{j,\alpha}(s),s\right),
\end{equation}
with 
\begin{equation}
\label{eq:Deltan}
\begin{aligned}
\Delta^n\left(h'_{j,\alpha}(s),s\right) &= \int_{-\infty}^{\infty} \dd{t} W_\lambda(t)\,
\left[ \tau_t^{H_{j,\alpha+n}(s)} - \tau_t^{H_{j,\alpha+n-1}(s)} \right]\left(h'_{j,\alpha}(s)\right), \quad n>0, \\
\Delta^0\left(h'_{j,\alpha}(s),s\right) &= \int_{-\infty}^{\infty} \dd{t} W_\lambda(t)\,
\tau_t^{H_{j,\alpha}(s)}\left(h'_{j,\alpha}(s)\right).
\end{aligned}
\end{equation}
Where $H_{j,\alpha}(s)$ is the Hamiltonian restricted to the terms which are fully supported in $B_{j,\alpha}$.
\begin{equation}
H_{j,\alpha}(s) = \sum_{\substack{i,\beta:\\B_{i,\beta}\subseteq B_{j,\alpha}}}h_{i,\beta}(s).
\end{equation}
Notice that $\supp\left[\Delta^n\left(h'_{j,\alpha}(s),s\right)\right] = B_{j,\alpha+n}$.
We now rewrite eq.~\eqref{eq:quasiadiabatic Delta} as a sum over all balls $B_{i,\alpha}$ of terms that are supported in $B_{i,\alpha}$,
\begin{equation}
\label{eq:kjalpha def}
\begin{split}
K(s) &= \sum_{j,\alpha}\sum_{n\geq 0} \sum_{\substack{i,\beta:\\B_{i,\beta+n}=B_{j,\alpha}}} \Delta^n\left(h'_{j,\beta}(s),s\right) \\
&= \sum_{j,\alpha}\sum_{\beta=1}^{\alpha}\Delta^{\alpha-\beta}\left(h'_{j,\beta}(s),s\right) \overset{\text{def}}{\equiv} \sum_{j,\alpha}k_{j,\alpha}(s),
\end{split}
\end{equation}
where we defined the quasi-local terms $k_{j,\alpha}(s)=\sum_{\beta=1}^{\alpha}\Delta^{\alpha-\beta}\big(h'_{j,\beta}(s),s\big)$ and it is clear that $\supp\left[k_{j,\alpha}(s)\right]=B_{j,\alpha}$.

We still need to argue that the norm of these terms decays fast with the size of the support.
We proceed by showing that this decay is faster than any polynomial.
We need to control $\norm*{\Delta^n\big(h'_{j,\beta}(s),s\big)}$ and therefore $\norm*{\big[ \tau_t^{H_{j,\beta+n}(s)} - \tau_t^{H_{j,\beta+n-1}(s)} \big]\big(h'_{j,\beta}(s)\big)}$.
From now on, we will remove the $s$-dependence, since all the bounds will not depend on $s$.
Let's rewrite the last term as follows
\begin{equation}
\label{eq:Delta bound}
\begin{split}
&\norm{\left[ \tau_t^{H_{j,\beta+n}} - \tau_t^{H_{j,\beta+n-1}} \right]\left(h'_{j,\beta}\right)} \\
&\qquad\qquad= \norm{\int_0^t \dd{t'} \dv{t'}\, \tau_{t'}^{H_{j,\beta+n-1}} \circ \tau_{t-t'}^{H_{j,\beta+n}}\left(h'_{j,\beta}\right) } \\
&\qquad\qquad= \norm{\int_0^t \dd{t'} \tau_{t'}^{H_{B_{j,\beta+n-1}}} 
\left(\left[H_{j,\beta+n-1}-H_{j,\beta+n},\tau_{t-t'}^{H_{j,\beta+n}}\left(h'_{j,\beta}\right)\right]\right) } \\
&\qquad\qquad\leq \sum_{i\in B_{j,\beta+n}}\int_0^{\abs{t}} \dd{t'} \norm{\left[h_{i,\beta+n-d_{i,j}},\tau_{t'}^{H_{j,\beta+n}}\left(h'_{j,\beta}\right)\right]}.
\end{split}
\end{equation}
The last equality is obtained by noticing that the difference $H_{j,\beta+n-1}-H_{j,\beta+n}$ contains terms whose support may be centred anywhere in the ball $B_{j,\beta+n}$, but have to extend all the way to the boundary of $B_{j,\beta +n}$.
To bound the norm of the commutator in the last line of eq.~\eqref{eq:Delta bound} we use the Lieb-Robinson bounds that derive from the quasi-locality of $H(s)$.
Define $\bar d$ as the distance between the supports of the two local Hamiltonian terms, 
$\bar d=d(B_{j,\beta},B_{i,\beta+n-d_{i,j}})=\min\{2(d_{i,j}-\beta)-n,0\}$.
Then, let us apply eq.~\eqref{eq:Assumption4.4},
\begin{equation}
\label{eq:Delta bound 2}
\begin{split}
\sum_{i\in B_{j,\beta+n}}&\int_0^{\abs{t}} \dd{t'} \norm{\left[h_{i,\beta+n-d_{i,j}},\tau_{t'}^{H_{j,\beta+n}}\left(h'_{j,\beta}\right)\right]} \\
\lesssim& \sum_{i\in B_{j,\beta+n}}\int_0^{\abs{t}} \dd{t'} \norm{h_{i,\beta+n-d_{i,j}}}\norm{h'_{j,\beta}}
\min\bigg\{1,e^{a (v_a\abs*{t'}-\bar d)}\sum_{\substack{x\in B_{j,\beta}\\y\in B_{i,\beta+n-d_{i,j}}}}F(d_{x,y})\bigg\}.
\end{split}
\end{equation}
The sum in the last expression can be bounded by using the integrability condition eq.~\eqref{eq:F integrability},
\begin{equation}
\sum_{\substack{x\in B_{j,\beta}\\y\in B_{i,\beta+n-d_{i,j}}}}F(d_{x,y}) \leq \abs{B_{j,\beta}}\max_{x\in B_{j,\beta}} \sum_{y\in B_{i,\beta+n-d_{i,j}}}F(d_{x,y})
\leq \abs{B_{j,\beta}} \norm{F}.
\end{equation}
Moreover, the norms of the Hamiltonian terms are bounded by hypothesis (see discussion after eq.~\eqref{eq:F convolution}), therefore eq.~\eqref{eq:Delta bound 2} is bounded by
\begin{equation}
\lesssim\sum_{i\in B_{j,\beta+n}}\int_0^{\abs{t}} \dd{t'} F_a\left(2(\beta+n-d_{i,j})\right)\frac{F_b\left(2\beta\right)}{\abs{B_{j,\beta}}}
\min\left\{1,e^{a (v_a\abs*{t'}-\bar d)}\abs{B_{j,\beta}}\right\}.
\end{equation}
To deal with the sum over all sites in $B_{j,\beta+n}$ we bound it with a sum over $\alpha=d_{i,j}=1,\dots,\beta+n$, and consider that for fixed $\alpha$ there are $\sim \alpha^{d-1}$ terms where the site $i$ is at distance $\alpha$ from site $j$.
These terms are all equal since the only dependence on $i$ is inside $d_{i,j}$.
Using also that $\abs{B_{j,\beta}}\sim \beta^d$,
\begin{equation}
\label{eq:sum over distances}
\lesssim\sum_{\alpha=1}^{\beta+n}\alpha^{d-1}
\int_0^{\abs{t}} \dd{t'} F_a\left(2(\beta+n-\alpha)\right)\frac{F_b\left(2\beta\right)}{\beta^d}
\min\left\{1,e^{a (v_a\abs*{t'}-\bar d)}\beta^d\right\}.
\end{equation}

The strategy is now to split the sum in two parts.
In the first part, for low values of $\alpha$, we will use the fact that the support of $h_{i,\beta+n-\alpha}$ is large and therefore $\norm{h_{i,\beta+n\alpha}}\lesssim F_a\left(2(\beta+n-\alpha)\right)$ is very small.
In the second part, for large $\alpha$, we will use the fact that the support of $h_{i,\beta+n-\alpha}$ and $h_{j,\beta}$ are far apart, and therefore the term coming from the Lieb-Robinson bound will be very small.
We split the sum at $\bar d=n/2$, which implies $\alpha=\beta+3/4n$.
In the first part of the sum we pick the constant term in the minimum coming from the Lieb-Robinson bound,
\begin{equation}
\label{eq:Delta bound 3.1}
\begin{split}
\sum_{\alpha=1}^{\beta+3/4n}&\alpha^{d-1}\int_0^{\abs{t}} \dd{t'} F_a\left(2(\beta+n-\alpha)\right)\frac{F_b\left(2\beta\right)}{\beta^d} \\
&= \sum_{\alpha=1}^{\beta+3/4n}\left(\frac{\alpha}{\beta}\right)^d\frac{\abs{t}}{\alpha}F_a\left(2(\beta+n-\alpha)\right)F_b\left(2\beta\right)\\
&\lesssim \left(1+\frac{3n}{4\beta}\right)^d\abs{t} F_b(2\beta)\sum_{\alpha=1}^{\beta+3/4n}F_a\left(2(\beta+n-\alpha)\right)\\
&\lesssim \left(1+\frac{3n}{4\beta}\right)^d\abs{t} F_b(2\beta) e^{-a\frac{n}{2}}.
\end{split}
\end{equation}
In the third line we roughly bounded the sum by substituting the polynomial terms that depend on $\alpha$ with the largest possible value.
In the last line, we simply used the definition of $F_a$, pulled out of the sum the exponential part evaluated in the worst case, $\alpha=\beta+3/4n$, and used the integrability condition of $F$, eq.~\eqref{eq:F integrability}.
In the second part of the sum in $\alpha$, the exponential decay is guaranteed by the Lieb-Robinson bound,
\begin{equation}
\label{eq:Delta bound 3.2}
\begin{split}
\sum_{\alpha=\beta+3/4n}^{\beta+n}&\alpha^{d-1}\int_0^{\abs{t}} \dd{t'} F_a\left(2(\beta+n-\alpha)\right)F_b\left(2\beta\right)e^{a (v_a\abs*{t'}-\bar d)} \\
&\lesssim e^{a v_a\abs*{t}}F_b\left(2\beta\right)\sum_{\alpha=\beta+3/4n}^{\beta+n}\alpha^{d-1}F_a\left(2(\beta+n-\alpha)\right)e^{-a\bar d}\\
&\lesssim e^{a v_a\abs*{t}}\left(\beta+n\right)^{d-1}F_a\left(0\right)F_b\left(2\beta\right)\sum_{\alpha=\beta+3/4n}^{\beta+n}e^{-a\left(2(\alpha-\beta)-n\right)}\\
&\lesssim e^{a v_a\abs*{t}}\left(\beta+n\right)^{d-1}F_a\left(0\right)F_b\left(2\beta\right)\frac{n}{4}e^{-a\frac{n}{2}}.
\end{split}
\end{equation}
Again we roughly bounded the sum by taking the largest term times the number of terms in the sum.
These very rough estimates can of course be refined, but this is already enough to our purposes.
Finally, using this estimates we can bound eq.~\eqref{eq:Delta bound}.
For large $n$, the second term eq.~\eqref{eq:Delta bound 3.2} will be larger than the one in eq.~\eqref{eq:Delta bound 3.1}.
\begin{equation}
\norm{\left[ \tau_t^{H_{j,\beta+n}} - \tau_t^{H_{j,\beta+n-1}} \right]\left(h'_{j,\beta}(s)\right)}
\lesssim n \left(\beta+n\right)^{d-1}F_b\left(2\beta\right)e^{-a\left(\frac{n}{2}-v_a \abs{t}\right)}.
\end{equation}

Using this last result we can bound $\Delta^n\left(h'_{j,\alpha}(s),s\right)$ in eq.~\eqref{eq:Deltan}.
To do so, we need to split the integral and use the properties of the filter function $W_\lambda(t)$ for $t\to\infty$
\begin{equation}
\begin{split}
&\norm{\Delta^n\left(h'_{j,\alpha}(s),s\right)}
\leq\int_{-\infty}^{\infty} \dd{t} \abs{W_\lambda(t)}\,\norm{\left[ \tau_t^{H_{j,\alpha+n}(s)} - \tau_t^{H_{j,\alpha+n-1}(s)} \right]\left(h'_{j,\alpha}(s)\right)} \\
&\lesssim \int_{-T}^{T} \dd{t} \abs{W_\lambda(t)}\,n \left(\beta+n\right)^{d-1}F_b\left(2\beta\right)e^{-a\left(\frac{n}{2}-v_a \abs{t}\right)}
+ 2\int_{|t|>T} \dd{t} \abs{W_\lambda(t)}\,\norm{h'_{j,\beta}(s)}.
\end{split}
\end{equation}
Using that $\norm*{h'_{j,\beta}(s)}\lesssim F_b(2\beta)$ and the properties of integrability at infinity of $W_\lambda(t)$ (see Lemma~2.6 of \cite{Bachmann2012}),
\begin{equation}
\label{eq:Delta bound T}
\norm{\Delta^n\left(h'_{j,\alpha}(s),s\right)} \lesssim
\left(\beta+n\right)^{d-1}F_b\left(2\beta\right)e^{-a\left(\frac{n}{2}-v_a T\right)}
+ F_b\left(2\beta\right)I_\lambda(T).
\end{equation}
The function $I_\lambda(t)$ is shown in Ref.~\cite{Bachmann2012} to be bounded for large $t$ by $I_\lambda(t)\lesssim (\lambda t)^{10} u_{2/7}(\lambda t)$, where $u_\mu(t)=e^{-\mu\frac{t}{\log^2(t)}}$.
Finally, choosing $T=n/(4v_a)$, we get the desired bound
\begin{equation}
\label{eq:Delta bound final}
\norm{\Delta^n\left(h'_{j,\alpha}(s),s\right)} \lesssim
\left(\beta+n\right)^{d-1}F_b\left(2\beta\right)e^{-a\frac{n}{4}}
+ F_b\left(2\beta\right)I_\lambda\left(\frac{n}{4v_a}\right).
\end{equation}
The function $I_\lambda$ vanishes faster than any polynomial but slower than exponential, therefore it is the largest term in eq.~\eqref{eq:Delta bound final} for large $n$.

With the last estimate we can finally bound the terms $k_{j,\alpha}(s)$ defined in eq.~\eqref{eq:kjalpha def}
\begin{equation}
\begin{split}
\norm{k_{j,\alpha}(s)} &\leq \sum_{\beta=1}^{\alpha}\norm{\Delta^{\alpha-\beta}\left(h'_{j,\beta}(s),s\right)}\\
&\lesssim \sum_{\beta=1}^{\alpha} F_b\left(2\beta\right)I_\lambda\left(\frac{\alpha-\beta}{4v_a}\right)
\lesssim \sum_{\beta=1}^{\alpha} e^{-2b(\alpha-\beta)} I_\lambda\left(\frac{\beta}{4v_a}\right),
\end{split}
\end{equation}
where we used that $F(x)\leq F(0)$.
To estimate the last sum we again split it in two parts:
in the first, for small $\beta$, there is an exponential decay, while for large $\beta$ the function $I_\lambda$ will vanish
\begin{equation}
\label{eq:kja norm}
\begin{split}
\norm{k_{j,\alpha}(s)}
&\lesssim \sum_{\beta=1}^{\alpha/2} e^{-2b(\alpha-\beta)} I_\lambda\left(\frac{\beta}{4v_a}\right)
+\sum_{\beta=\alpha/2}^{\alpha} e^{-2b(\alpha-\beta)} I_\lambda\left(\frac{\beta}{4v_a}\right)\\
&\lesssim \frac{\alpha}{2} I_\lambda(1)e^{-2b\frac{\alpha}{2}} + \frac{\alpha}{2}I_\lambda\left(\frac{\alpha}{8v_a}\right)
\lesssim \alpha I_\lambda\left(\frac{\alpha}{8v_a}\right).
\end{split}
\end{equation}
In the second inequality we bounded the sums by the largest value for each factor, times the number of terms in the sum, while in the third we kept only the most relevant term for large $\alpha$.
With this we conclude the proof of the quasi-locality of the quasi-adiabatic Hamiltonian $K(s)$.
Actually, we did not prove that it satisfies eq.~\eqref{eq:quasilocality} for some $F$.
This is done in Ref.~\cite{Bachmann2012}, however the result~\eqref{eq:kja norm} is what we need for what comes next.

\subsubsection{Approximating the quasi-adiabatic evolution with a finite depth quantum circuit}
\label{subsubsec:K finite depth circuit}

In this section we will adapt the proof of Ref.~\cite{Osborne2005} to the case of the quasi-adiabatic evolution.
The main difference is that the quasi-adiabatic Hamiltonian $K(s)$ is not strictly local, but quasi-local.
With respect to Ref.~\cite{Osborne2005}, we will also consider the case of general dimensionality $d$ of the system.

\begin{substep}
For $t\in [0,1]$, the quasi-adiabatic evolution operator $\exp\left[\iu \int_0^t \dd{s}K(s)\right]$ can be approximated with a finite depth quantum circuit $C$ with gates of diameter $\Omega$, with an error given by the following bound:
\begin{equation}
\norm{\exp\left[\iu \int_0^t \dd{s}K(s)\right] - C} \lesssim \poly(N,\Omega)\, \tilde u_\mu\left(b\,\Omega\right),
\end{equation}
where $b$ is a constant and $\tilde u_\mu(\eta) = u_\mu(\eta)$ for $\eta\geq e^2$ and a constant $\tilde u_\mu(\eta)=u_\mu(e^2)$ otherwise.
In particular, for $\Omega\sim \log^{1+\delta} N$ with $\delta>0$, the bound vanishes as $N\to\infty$ faster than any power.
\end{substep}

To start, consider a simply connected region $A$ in the lattice.
Consider then the Hamiltonian formed by the terms of the quasi-adiabatic Hamiltonian that are fully supported in $A$,
\begin{equation}
\label{eq:K_A def}
K_A(s)=\sum_{\substack{j,\alpha:\\B_{j,\alpha}\subseteq A}} k_{j,\alpha}(s).
\end{equation}
We ultimately want to approximate the quasi-adiabatic evolution $\exp\left[\iu \int_0^t \dd{s}K(s)\right]$ for $t\in[0,1]$ with a quantum circuit, namely a product of unitary operators with finite support (actually polylogarithmic in the system size).
Consider first the product of two evolution operators that are fully localized in $A$ and its complement $\compl A$, namely $\exp\left[\iu \int_0^t \dd{s}K_A(s)\right]\otimes \exp\left[\iu \int_0^t \dd{s}K_{\compl A}(s)\right]$.
Approximating $\exp\left[\iu \int_0^t \dd{s}K(s)\right]$ with this operator, completely disregarding all terms in $K(s)$ whose support has non-zero intersection with the boundary of $A$ (which we denote by $\partial A$) would be a very crude approximation.
We then need to consider the ``patch'' operator
\begin{equation}
V(t)=\left(e^{-\iu \int_0^t \dd{s}K_A(s)}\otimes e^{-\iu \int_0^t \dd{s}K_{\compl A}(s)}\right)e^{\iu \int_0^t \dd{s}K(s)}.
\end{equation}
This operator is supported on the full lattice and in order to obtain a quantum circuit we need to approximate it with some $V_\Omega(t)$ that acts on a finite number of sites.
First of all we notice that $V(t)$ satisfies the following differential equation
\begin{equation}
\dv{V}{t} (t) = \iu\, V(t)\,\alpha_t\Big(K(t)-K_A(t)-K_{\compl A}(t)\Big),
\end{equation}
which is solved by
\begin{equation}
V(t) = \exp\left[\iu\int_0^t \dd{s} L(s)\right],
\end{equation}
with generator
\begin{equation}
L(t) \overset{\text{def}}{\equiv} \alpha_t\Big(K(t)-K_A(t)-K_{\compl A}(t)\Big) = \alpha_t\left(\sum_{B_{j,\alpha}\cap\partial A\neq\emptyset}k_{j,\alpha}(t)\right).
\end{equation}
We will show that the generator of $V(t)$ is quasi-local.
It can be written as a sum of terms $L(t) = \sum_{j,\alpha}\ell_{j,\alpha}$, with norm $\norm{\ell_{j,\alpha}}$ decreasing faster than any polynomial in the linear size of the support $\alpha$.
Using Lieb-Robinson bounds, we then show that only the terms fully supported in balls close to $\partial A$ are relevant and therefore $L(t)$ can be efficiently approximated with an operator $L_\Omega (t)$ which acts on sites at most at distance $\Omega$ from $\partial A$.
The parameter $\Omega$ will determine the size of the support of $L_\Omega(t)$ (at least in one of the directions of the lattice) and ultimately the size of the gates of the circuit.
Defining $V_\Omega(t)$ through $\dv*{V_\Omega(t)}{t} = V_\Omega(t) L_\Omega(t)$, and integrating $\norm{V(t)-V_\Omega(t)}\leq \int_0^{\abs{t}}\norm{L(t)-L_\Omega(t)}$, we get a bound on the error that we make by using the local unitary operator $V_\Omega(t)$ in place of $V(t)$.
To obtain the circuit we will proceed by choosing some $A$, and approximating the evolution by two gates localized in $A$ and $\compl A$ and a third patch gate $V_\Omega(t)$ that acts on another level of the circuit.
Some or all of these gates could still be of size $\sim N$ in one or more directions of the lattice.
We will then repeat this approximation scheme until we obtain a circuit with all gates of maximum linear size $\Omega$.
There are many geometrical schemes to do this, depending on the shape of the lattice and in particular on the dimensionality of the system.
The depth of the final circuit will depend on the scheme used and in general on the dimensionality of the lattice.
However, one can always choose a scheme to get a circuit with finite depth.
Notice also that $L(s)$ has the same quasi-local structure of $K(s)$, and therefore the same approximation scheme can be performed on $L(s)$ or $L_\Omega(s)$.
We will find that to have a good approximation of the quasi-adiabatic evolution we have to choose $\Omega\sim \log^{1+\delta}N$ for any $\delta>0$.
Therefore the gates are not strictly local, however this $\poly\log N$ scaling is enough to ultimately build a $\log$-local Lindbladian $\mathcal L$ which satisfies some Lieb-Robinson bounds and such that $\rho_0\fastto[\mathcal L]\rho_1$.

Let's start by showing that $L(t)$ is quasi-local.
First we write it as a sum of terms supported on the balls of the lattice, in the same way as we did for $K(s)$ in Section~\ref{subsubsec:K quasi-local},
\begin{equation}
\label{eq:L Delta}
L(t) = \sum_{B_{j,\alpha}\cap\partial A\neq\emptyset} \alpha_t\left(k_{j,\alpha}(t)\right)
 = \sum_{B_{j,\alpha}\cap\partial A\neq\emptyset}\sum_{n=0}^{\infty}\Delta^n(k_{j,\alpha}(t),t) ,
\end{equation}
where $\Delta^n(k_{j,\alpha}(t),t)=\big(\alpha_t^{j,\alpha+n}-\alpha_t^{j,\alpha+n-1}\big)\left(k_{j,\alpha}(t)\right)$ for $n>0$, while for $n=0$ we define
$\Delta^0(k_{j,\alpha}(t),t)=\alpha_t^{j,\alpha}\left(k_{j,\alpha}(t)\right)$.
Here $\alpha_t^{j,\alpha}$ is the  superoperator implementing the evolution according to $K_{B_{j,\alpha}}(s)$.
Notice that the supports of the terms in eq.~\eqref{eq:L Delta} are the balls $B_{j,\alpha+n}$ and have non zero intersection with $\partial A$.
Then, for any $B_{j,\alpha}$ intersecting $\partial A$ we gather all terms supported on the balls with centre in $j$ and radius smaller than $\alpha$ (and still non zero intersection with $\partial A$),
\begin{equation}
L(t) = \sum_{B_{j,\alpha}\cap\partial A\neq\emptyset} \sum_{\substack{\beta=1\\B_{j,\beta}\cap\partial A\neq\emptyset}}^{\alpha}
\Delta^{\alpha-\beta}\left(k_{j,\beta}(t),t\right)
\overset{\text{def}}{\equiv} \sum_{j,\alpha}\ell_{j,\alpha}(t),
\end{equation}
where we defined the $\ell_{j,\alpha}(t)$'s, with $\supp\left[\ell_{j,\alpha}(t)\right]=B_{j,\alpha}$.
Notice that the sum over $\beta$ in the definition of $\ell_{j,\alpha}$ can be rewritten as follows,
\begin{equation}
\label{eq:ell def}
\ell_{j,\alpha}(t) = \sum_{\beta=d(j,\partial A)}^{\alpha}\Delta^{\alpha-\beta}\left(k_{j,\beta}(t),t\right).
\end{equation}

We want to prove that $L(t)$ is quasi-local, and therefore we need to prove that $\norm{\ell_{j,\alpha}(t)}$ decays fast with $\alpha$.
As done with $K(s)$, let us first focus on $\Delta^n(k_{j,\beta}(t),t)$
\begin{equation}
\begin{split}
\norm{\Delta^n\left(k_{j,\beta}(t),t\right)} &= \norm{\big(\alpha_t^{j,\beta+n}-\alpha_t^{j,\beta+n-1}\big)\left(k_{j,\beta}(t)\right)}\\
&\leq \int_0^{\abs{t}}\dd{t'} \norm{\left[K_{B_{j,\beta+n}}(t')-K_{B_{j,\beta+n-1}}(t'),\alpha_{t':t}^{j,\beta+n}\left(k_{j,\beta}(t)\right)\right]}\\
&\leq \sum_{i\in B_{j,\beta+n}} \int_0^{\abs{t}}\dd{t'} \norm{\left[k_{i,\beta+n-d_{i,j}}(t'),\alpha_{t':t}^{j,\beta+n}\left(k_{j,\beta}(t)\right)\right]}.
\end{split}
\end{equation}
Here $\alpha_{t':t}^{j,\beta}$ denotes the evolution according to $K_{B_{j,\beta}}(s)$ in the time interval $[t',t]$.
The last inequality can be understood as before by noticing that the support of the terms that do not cancel out can be centred anywhere in $B_{j,\beta+n}$ but have to extend up to the boundary of $B_{j,\beta+n}$.
We can now use Lieb-Robinson bound for $\alpha_t$ that have been proven in Ref.~\cite{Bachmann2012}, Theorem~4.5,
\begin{equation}
\begin{split}
&\norm{\Delta^n\left(k_{j,\beta}(t),t\right)} 
\lesssim \sum_{i\in B_{j,\beta+n}} \int_0^{\abs{t}}\dd{t'} \norm{k_{i,\beta+n-d_{i,j}}(t')}\norm{k_{j,\beta}(t)}\times\\
&\times\min\bigg\{1,e^{v t'}\abs{B_{j,\beta}}\tilde u_\mu\left(\frac{\lambda}{8v_a}d\left(B_{j,\beta},B_{i,\beta+n-d_{i,j}}\right)\right)\max_{x\in B_{i,\beta}}\sum_{y\in B_{i,\beta+n-d_{i,j}}}\!\!\!\!\!\! F\left(\frac{\lambda}{8v_a}d_{x,y}\right)\bigg\},
\end{split}
\end{equation}
where $\tilde u_\mu(\eta) = u_\mu(\eta)$ for $\eta\geq e^2$ and a constant $\tilde u_\mu(\eta)=u_\mu(e^2)$ otherwise.
Here $0<\mu<2/7$.
We can discard the last term using the integrability of $F$, eq.~\eqref{eq:F integrability}.
We define $\bar d =\lambda/(8v_a)\min\{0,2(d_{i,j}-\beta)-n\}$, and after using the bound \eqref{eq:kja norm} we can perform the integration in time.
As done for $K(s)$ in eq.~\eqref{eq:sum over distances}, we can then rewrite the sum over $\alpha=d_{i,j}$ and remember that there are $\sim\alpha^{d-1}$ balls centred at a distance $\alpha$ from site $j$ and with fixed radius $\beta+n-\alpha$,
\begin{equation}
\label{eq:Delta bound L}
\begin{split}
&\norm{\Delta^n\left(k_{j,\beta}(t),t\right)} \\
&\phantom{\||}\lesssim \sum_{i\in B_{j,\beta+n}} (\beta+n-d_{i,j})I_\lambda\left(\frac{\beta+n-d_{i,j}}{8v_a}\right)\beta I_\lambda\left(\frac{\beta}{8v_a}\right)
\min\left\{t,e^{vt}\abs{B_{j,\beta}}\tilde u_\mu(\bar d)\right\}\\
&\phantom{\||}\lesssim \sum_{\alpha=1}^{\beta+n}\alpha^{d-1}(\beta+n-\alpha)I_\lambda\left(\frac{\beta+n-\alpha}{8v_a}\right)\beta I_\lambda\left(\frac{\beta}{8v_a}\right)
\min\left\{t,e^{vt}\abs{B_{j,\beta}}\tilde u_\mu(\bar d)\right\}.
\end{split}
\end{equation}
We again split the sum in two parts, $1\leq\alpha\leq \beta+3/4n$ and $\beta+3/4n\leq\alpha\leq \beta+n$.
The first part will be small due to the fact that the support of $k_{i,\beta+n-\alpha}$ is large,
\begin{equation}
\label{eq:smaller alpha}
\begin{split}
\sum_{\alpha=0}^{\beta+3/4n}&\alpha^{d-1} (\beta+n-\alpha)\beta I_\lambda\left(\frac{\beta+n-\alpha}{8v_a}\right) I_\lambda\left(\frac{\beta}{8v_a}\right)\\
&\lesssim \beta^M u_{2/7}\left(\frac{\lambda}{8v_a}\beta\right)\sum_{\alpha=0}^{\beta+3/4n}\alpha^{d-1}(\beta+n-\alpha)^M u_{2/7}\left(\frac{\lambda}{8v_a}(\beta+n-\alpha)\right)\\
&\lesssim \beta^M (\beta+n)^M \left(\beta+\frac{3}{4}n\right)^{d-1} u_{2/7}\left(\frac{\lambda}{8v_a}\beta\right)u_{2/7}\left(\frac{\lambda}{32v_a}n\right),
\end{split}
\end{equation}
for some integer $M$.
The second part of the sum will vanish thanks to the Lieb-Robinson bound,
\begin{equation}
\label{eq:larger alpha}
\begin{split}
e^{vt}&\sum_{\alpha=\beta+3/4n}^{\beta+n} \alpha^{d-1} (\beta+n-\alpha)^M\beta^{d+M} \times \\
&\qquad\times u_{2/7}\left(\frac{\lambda}{8v_a}\beta\right)
u_{2/7}\left(\frac{\lambda}{8v_a}(\beta+n-\alpha)\right)
\tilde u_\mu\left(\frac{\lambda}{8v_a}[2(\alpha-\beta)-n]\right)\\
&\lesssim e^{vt}\left(\frac{n}{4}\right)^{M+1}\beta^{d+M}(\beta+n)^{d-1}u_{2/7}\left(\frac{\lambda}{8v_a}\beta\right)
\tilde u_\mu\left(\frac{\lambda}{16v_a}n\right).
\end{split}
\end{equation}
Using eqs.~\eqref{eq:smaller alpha} and~\eqref{eq:larger alpha} we can bound eq.~\eqref{eq:Delta bound L}.
Apart from constants in front of each term,
\begin{equation}
\begin{split}
\norm{\Delta^n\left(k_{j,\beta}(t),t\right)} 
\lesssim& \beta^M (\beta+n)^M \left(\beta+\frac{3}{4}n\right)^{d-1} u_{2/7}\left(\frac{\lambda}{8v_a}\beta\right)u_{2/7}\left(\frac{\lambda}{32v_a}n\right)+\\
&+e^{vt}n^{M+1}\beta^{d+M}(\beta+n)^{d-1}u_{2/7}\left(\frac{\lambda}{8v_a}\beta\right)
\tilde u_\mu\left(\frac{\lambda}{16v_a}n\right)\\
\lesssim& \poly(\beta,n)\,
u_{2/7}\left(\frac{\lambda}{8v_a}\beta\right)
\tilde u_\mu\left(\frac{\lambda}{16v_a}n\right).
\end{split}
\end{equation}
With this result we can bound the quasi-local terms in $L(t)$,
\begin{equation}
\begin{split}
\norm{\ell_{j,\alpha}(t)} &\leq \sum_{\beta=d(j,\partial A)}^{\alpha} \norm{\Delta^{\alpha-\beta}\left(k_{j,\beta}(t),t\right)}\\
&\lesssim \sum_{\beta=d(j,\partial A)}^{\alpha} \poly(\alpha,\beta) 
u_{2/7}\left(\frac{\lambda}{8v_a}\beta\right) \tilde u_\mu\left(\frac{\lambda}{16v_a}(\alpha-\beta)\right)\\
&\lesssim \poly(\alpha)\, \tilde u_\mu\left(\frac{\lambda}{16v_a}\alpha\right).
\end{split}
\end{equation}
To obtain the last inequality one has to use the property $u_\mu(x+y)\geq u_\mu(x)u_\mu(y)$, or the convolution property, eq.~\eqref{eq:F convolution}.

Now that we showed that $L(t)$ is quasi-local, we can define $L_\Omega(t)$ by keeping only the terms that are fully supported in the ``fattening'' of $\partial A$ by $\Omega$, namely $\partial A_\Omega=\{i\in\Lambda : d(i,\partial A)<\Omega\}$,
\begin{equation}
L_\Omega(t) = \sum_{\substack{B_{j,\alpha}\cap\partial A\neq\emptyset\\B_{j,\alpha}\subset \partial A_\Omega}}\ell_{j,\alpha}(t).
\end{equation}
It is now immediate to check that the difference between $L(t)$ and $L_\Omega(t)$ vanishes for large $N$ if $\Omega$ is chosen to scale as $\Omega\sim\log^{1+\delta}N$ for any $\delta>0$.
Indeed, the difference contains only terms whose support is large, since they must have non zero intersection with $\partial A$ and are not fully contained in $\partial A_\Omega$.
Therefore their size must be at least of order $\Omega$.
We can actually find a very crude bound for the difference by considering all terms with a support with a linear size larger than $\sim\Omega$.
The very fast decay of their norm will guarantee that the bound will vanish for large $N$.
More precisely,
\begin{equation}
\label{eq:L approx}
\begin{split}
\norm{L(t)-L_\Omega(t)}
&= \Big\lVert\sum_{\substack{B_{j,\alpha}\cap\partial A\neq\emptyset\\B_{j,\alpha}\nsubseteq \partial A_\Omega}}\ell_{j,\alpha}(t)\Big\rVert
\leq \sum_{\substack{B_{j,\alpha}\cap\partial A\neq\emptyset\\B_{j,\alpha}\nsubseteq \partial A_\Omega}}\norm{\ell_{j,\alpha}(t)} \\
&\lesssim \sum_{j\in\Lambda} \sum_{\alpha\geq\Omega/2}\norm{\ell_{j,\alpha}(t)}
\lesssim N\sum_{\alpha \geq \Omega/2}\poly(\alpha) \, \tilde u_\mu\left(\frac{\lambda}{16v_a}\alpha\right)\\
&\lesssim \poly(N,\Omega)\, \tilde u_\mu\left(\frac{\lambda}{32v_a}\Omega\right).
\end{split}
\end{equation}
The last inequality can be obtained by noticing that the function $\poly(\alpha)\,u_\mu(\alpha)$ is decreasing for large $\alpha$, the largest term in the sum is the one for $\alpha=\Omega/2$ and the sum has less than $N^{1/d}$ terms.
Another way is to use the integration property of $u_\mu(x)$ discussed in Lemma~2.5 of Ref.~\cite{Bachmann2012}.
From eq.~\eqref{eq:L approx} we see that if indeed we choose $\Omega$ to grow faster than logarithmic in the system size, then the error that we do by approximating $L(t)$ with $L_\Omega(t)$ decays to zero with $N$ faster than any polynomial.

Let us summarize the result we obtained up to here:
given a region $A\subset\Lambda$, we showed that we can approximate the quasi adiabatic evolution operator $\exp\left[\iu \int_0^t \dd{s}K(s)\right]$ with the product of the disentangled operator $\exp\left[\iu \int_0^t \dd{s}K_A(s)\right]\otimes \exp\left[\iu \int_0^t \dd{s}K_{\compl A}(s)\right]$ times the unitary ``patch'' operator $V_\Omega(s)$, which is localized around the boundary between $A$ and $\compl A$, $\supp\left[V_\Omega(t)\right]=\partial A_\Omega$.
Moreover it has the same quasi-local structure of $K(s)$.
If we choose $\Omega\sim\log^{1+\delta}N$, the error of this approximation vanishes faster than any polynomial in $N$.
By iterating this procedure we can approximate the operator $\exp\left[\iu \int_0^t \dd{s}K(s)\right]$ with a unitary quantum circuit with finite depth.
During this procedure, it may be necessary to apply the same approximation scheme also to some patches $V_\Omega$'s in case the support $\partial A_\Omega$ extends through the whole lattice in some directions.
This can be done, since the structure of quasi-locality of $L(t)$ is the same as the one of $K(t)$ and therefore satisfies the same Lieb-Robinson bounds.

Let us explicitly describe this approximation scheme in the case of a two-dimensional square lattice.
One way to do it is to divide the lattice in strips $A_i$ of width $\Omega$ in the $x$ direction and length $\sqrt N$ in the $y$ direction.
The first layer of the circuit will be $e^{\iu \int_0^t \dd{s}K_{A_1}(s)}\otimes e^{\iu \int_0^t \dd{s}K_{A_2}(s)}\otimes\dots$.
The second layer consists in the patch operators which are supported around the boundary between $A_1$ and $A_2$, between $A_2$ and $A_3$ and so on.
They too, are of size $\Omega$ in the $x$ direction and $\sqrt N$ in the $y$ direction.
All the gates are not local yet, since they act on regions of sites that extend through the whole lattice in the $y$ direction.
For each of them we proceed with another set of approximations, by dividing it in regions of size $\Omega\times\Omega$ and add another set of patch operators, this time of size $\Omega\times\Omega$.
We have obtained the desired result, a circuit of unitary gates that act on a set of sites of size $\poly\log N$.
We need to do this second round of approximations for both layers of the previous step, and the number of layers is doubled.
Therefore the final $\log$-local circuit has depth four.
It is clear that this procedure can be extended to any dimensionality of the square lattice, and the depth of the circuit will be $2d$.

\subsection{Fast Lindbladian evolution from a finite depth circuit}
\label{subsec:L from circuit}

In this section we will build a time independent Lindbladian that drives the state $\rho_0=\ketbra{\psi_0}$ to $\rho_1=\ketbra{\psi_1}$ in $\poly\log$-time by effectively implementing the finite depth quantum circuit introduced in Section~\ref{subsec:finite depth circuit}.
We will make use of the same ancillary timers of Section~\ref{subsec:transitivity}.

\begin{step}
Given two pure states $\rho_0$ and $\rho_1$, and a finite depth quantum circuit $C$ such that $\rho_1=C\rho_0C^\dag$, there exists a Lindbladian $\mathcal L$ acting on the original system and a locality preserving ancillary system (as defined in observation~\ref{it:LR bounds} in Section~\ref{subsec:formal def}) such that for any $t\gtrsim \poly\log N$
\begin{equation}
\norm{e^{t\mathcal L}(\rho_0\otimes \phi_0) - \rho_1\otimes \phi_T}_1 \lesssim \eps_N ,
\end{equation}
with $\phi_0$ and $\phi_T$ respectively the initial and final state of the ancillas, and $\eps_N$ vanishing for large $N$.
\end{step}

For every gates of the circuit let us define the Hamiltonian term $U^{(\ell)}_{i}=e^{\iu h_i^{(\ell)}}$, for $\ell=1,\dots,L$ the `layer' index and $i=1,\dots,M$ the `space' index within the layer $\ell$.
The number of gates $M$ within each layer is of order $M\sim N/\log^{1+\delta}N$, and the $h^{(\ell)}_i$'s can be chosen to have norm $\norm*{h^{(\ell)}_i}=1$.
We define $L$ commuting Hamiltonians $H^{(\ell)} = \sum_i h^{(\ell)}_i$ and their corresponding Lindblad superoperators $\mathcal L_\ell(\rho) = \iu \left[H^{(\ell)},\rho\right]$.
We then introduce $M$ timers, each attached to a spin in the intersection between the support of the Hamiltonian terms in different layers.
We couple the corresponding switches to the system in a way that for a time $\tau_1=1$ the system evolves according to $H^{(1)}$, then for a time $\tau_2=1$ evolves with $H^{(2)}$ and so on.
After the last Hamiltonian is applied for a time $\tau_L=1$, the final state of the timers should induce no more evolution on the system.
In the following for simplicity we will focus on the case with only two layers, since it will be clear from the strategy of the proof that it can be easily extended to any number of layers.
The only difference with the case of Section~\ref{subsec:transitivity} is what happens in the time interval between $\tau_1$ and $\tau_2$.
Once this is established, adding layers and therefore switching times $\tau_\ell$ will only require to repeat the same analysis for any time intervals $\tau_{\ell-1}\leq t \leq\tau_\ell$.

For the $i$-th terms of the Hamiltonians $H^{(\ell)}$ we need two switches that change the evolution from $h_i^{(1)}$ to $h_i^{(2)}$ (they are chosen such that their supports have non zero intersection) and then to $h_i^{\mathrm{final}}=0$.
Then for each term consider a timer made of $T+1$ qubits initialized in the state $\phi_0^{(i)}=\ket{1}\otimes\ket{0}^{\otimes T}$ that evolve according to the Lindbladian in eq.~\eqref{eq:timer evolution}.
For a general two-time timer the first switch will be the $T_1$-th qubit.
At time $\tau_1=T_1/\gamma$ it flips to $\ket{1}_{T_1}$ and $H^{(1)}$ is substituted with $H^{(2)}$, while when the final qubit is flipped, the Hamiltonian is turned off.
The Hamiltonian of the full system with ancillas is then $H = \sum_i h_i$ with 
$h_i = \ketbrasub{0}{0}{T_1}\otimes \mathbb{I}_T\otimes h_i^{(1)} + \ketbrasub{1}{1}{T_1}\otimes\ketbrasub{0}{0}{T}\otimes h_i^{(2)}$.
We also define $T_2=T-T_1$ and $\tau_2=T_2/\gamma$.
The dissipative evolution will run for a total time $\tau=\tau_1+\tau_2=T/\gamma$.
Notice that since $\tau_1=\tau_2=1$, then $T_1=T_2$ and they scale with $N$ in the same way as $T$.

Analogously to what was done to show transitivity, let us consider the time intervals defined by $\tau_1^\pm=\tau_1\pm\eps/2$ and $\tau^\pm=\tau\pm\eps/2$.
As before, $\eps$ is chosen as function of $T$ and scales has $\eps\sim T^{-\alpha}$ with $0<\alpha<1/2$.
For $t\leq\tau_1^-$ the state can be written in the same way as in eq.~\eqref{eq:before tau}
\begin{equation}
\label{eq:before tau1}
e^{t\mathcal L}(\phi_0\otimes\rho_0) = \sum_{k_1,\dots,k_N = 0}^{T_1-1} p_{\{k_i\}}(t)\,\phi_{\{k_i\}} \otimes e^{t\mathcal L_1}(\rho_0) 
+ \sideset{}{'}\sum_{\{k_i\}} p_{\{k_i\}}(t)\,\phi_{\{k_i\}}\otimes \rho_{\{k_i\}}',
\end{equation}
and now the primed sum means the sum over all configurations of $\{k_i\}_i$ with at least one $k_j\geq T_1$.
As before, this allows us to conclude that
\begin{equation}
\label{eq:before tau1 result}
\Tr_T\left[e^{\tau_1^-\mathcal L}(\phi_0\otimes\rho_0)\right] = e^{\tau_1^- \mathcal L_1}(\rho_0) + \err{\experr[M][1]}.
\end{equation}
The first $\eps$-window around $\tau_1$ can be treated using the same argument of Section~\ref{subsec:transitivity},
\begin{equation}
\label{eq:window tau1}
\Tr_T\left[e^{\tau_1^+ \mathcal L}(\phi_0\otimes\rho_0)\right] 
\simeq \Tr_T\left[e^{\tau_1^- \mathcal L}(\phi_0\otimes\rho_0)\right] + \err{\eps M}.
\end{equation}

Let us now move to the interval $\tau_1^+ \leq t \leq \tau^-$.
We claim that in this time interval all timers will be almost surely in a state $\phi_{\{k_i\}}$ with $T_1\leq k_i<T$.
To prove this we have to show that $\sum_{k_1,\dots, k_N = T_1}^{T-1}p_{\{k_i\}}(t) \to 1$ when $T\to\infty$.
Indeed,
\begin{equation}
\sum_{k_1,\dots, k_N = T_1}^{T-1}p_{\{k_i\}}(t) = \prod_{i=1}^{M} \sum_{k_i=T_1}^{T-1} p_{k_i}(t) = \left[e^{-t\gamma}\sum_{k=T_1}^{T-1}\frac{(t\gamma)^k}{k!}\right]^M.
\end{equation}
The factor in parenthesis can be rewritten as follows
\begin{equation}
\label{eq:prob intermediate interval}
e^{-t\gamma}\sum_{k=T_1}^{T-1}\frac{(t\gamma)^k}{k!} = 1 - e^{-t\gamma}\sum_{k=0}^{T_1-1}\frac{(t\gamma)^k}{k!} - e^{-t\gamma}\sum_{k=T}^{\infty}\frac{(t\gamma)^k}{k!},
\end{equation}
and both sums in the rhs can be shown to go to zero exponentially.
Using Stirling's approximation, we can rewrite the first sum up to $T_1-1$ as follows
\begin{equation}
e^{-\frac{t}{\tau_1}T_1}\sum_{k=0}^{T_1-1}\left(\frac{t}{\tau_1}\right)^k\frac{T_1^k}{k!} \simeq
\frac{e^{-T_1\left[t/\tau_1 - \log(t/\tau_1) - 1\right]}}{\sqrt{2\pi T_1}} \frac{\tau_1}{t}\sum_{k=0}^{T_1-1}\left(\frac{\tau_1}{t}\right)^k \frac{(T_1-1)!}{T_1^k(T_1-k-1)!}.
\end{equation}
As was done in Section~\ref{subsec:transitivity}, it is easy to see that the last ratio is smaller than one, therefore we can bound this term as follows
\begin{equation}
\begin{split}
e^{-\frac{t}{\tau_1}T_1}&\sum_{k=0}^{T_1-1}\left(\frac{t}{\tau_1}\right)^k\frac{T_1^k}{k!} 
\lesssim\frac{e^{-T_1\left[t/\tau_1 - \log(t/\tau_1) - 1\right]}}{\sqrt{2\pi T_1}} \frac{\tau_1}{t}\sum_{k=0}^{T_1-1}\left(\frac{\tau_1}{t}\right)^k \\
&=\frac{e^{-T_1\left[t/\tau_1 - \log(t/\tau_1) - 1\right]}}{\sqrt{2\pi T_1}} \frac{\tau_1}{t}\frac{1-(\tau_1/t)^{T_1}}{1-\tau_1/t}
\lesssim \frac{e^{-T_1\left[t/\tau_1 - \log(t/\tau_1) - 1\right]}}{\sqrt{2\pi T_1}\left(t/\tau_1-1\right)} \xrightarrow[T\to\infty]{t\geq\tau_1^+} 0.
\end{split}
\end{equation}
Analogously, for the sum for $k>T$ in eq.~\eqref{eq:prob intermediate interval},
\begin{equation}
\begin{split}
e^{-\frac{t}{\tau}T}&\sum_{k=T}^\infty\left(\frac{t}{\tau}\right)^k\frac{T^k}{k!}
\simeq \frac{e^{-T\left[t/\tau - \log(t/\tau) - 1\right]}}{\sqrt{2\pi T}} \sum_{k=0}^\infty\left(\frac{t}{\tau}\right)^k T^j\frac{T!}{(T+k)!} \\
&\lesssim\frac{e^{-T\left[t/\tau - \log(t/\tau) - 1\right]}}{\sqrt{2\pi T}} \sum_{k=0}^\infty\left(\frac{t}{\tau}\right)^k
=\frac{e^{-T\left[t/\tau - \log(t/\tau) - 1\right]}}{\sqrt{2\pi T}\left(1-t/\tau\right)} \xrightarrow[T\to\infty]{t\leq\tau^-} 0.
\end{split}
\end{equation}
This proves that for $\tau_1^+ \leq t \leq \tau^-$,
\begin{equation}
\label{eq:tau1+<t<tau prob}
e^{-t\gamma}\sum_{k=T_1}^{T-1}\frac{(t\gamma)^k}{k!} \simeq 1 + \err{\experr[M]}.
\end{equation}
In the last expression there should be two errors, the one displayed and another term of the same form but with $T$, $\tau$ replaced with $T_1$, $\tau_1$.
However, since the scaling in $N$ is the same in the two cases, we write only one of them.
Also, we wrote a time-independent expression for the error, by taking the one at $t=\tau^-$ (or $\tau_1^+$).
With this result, we can conclude that for $\tau_1^+ \leq t \leq \tau^-$ the state of the system can be written as follows,
\begin{equation}
\label{eq:tau1+<t<tau initial}
e^{t\mathcal L}(\phi_0\otimes\rho_0) = \sum_{k_1,\dots, k_N = T_1}^{T-1} p_{\{k_i\}}(t)\,\phi_{\{k_i\}}\otimes \rho_{\{k_i\}}(t) + \err{\experr[M]}.
\end{equation}
Specializing this to $t=\tau_1^+$ and using the previous result at $t=\tau_1^-$ eq.~\eqref{eq:before tau1 result}, and the result for the $\eps$-time window $\tau_1^\pm$ eq.~\eqref{eq:window tau1},
\begin{equation}
\label{eq:state tau1+}
\Tr_T\left[e^{\tau_1^+\mathcal L}(\phi_0\otimes\rho_0)\right] \simeq \sum_{k_1,\dots, k_N = T_1}^{T-1} p_{\{k_i\}}(\tau_1^+)\,\rho_{\{k_i\}}(\tau_1^+)\simeq
e^{\tau_1^- \mathcal L_1}(\rho_0) + \err{\eps M},
\end{equation}
where once again in the error we kept only the most dominant contribution in the thermodynamic limit.

Now comes the main difference with the case discussed in Section~\ref{subsec:transitivity}:
In the time interval $\tau_1^+ \leq t \leq \tau^-$ we expect the original system to stop evolving with $\mathcal L_1$ and evolve (almost) exclusively with $\mathcal L_2$.
At the same time, the timers will still evolve, they are not yet in the fixed point of $\mathcal L_T$.
We then expect that the following holds:
\begin{equation}
	\label{eq:tau1+<t<tau}
\begin{split}
e^{t\mathcal L}(\phi_0\otimes\rho_0) &= e^{(t-\tau_1^+)\mathcal L}e^{\tau_1^+\mathcal L}(\phi_0\otimes\rho_0) \\
&\simeq \sum_{k_1,\dots, k_N = T_1}^{T-1} p_{\{k_i\}}(\tau_1^+)\,e^{(t-\tau_1^+)\mathcal L_T}\left(\phi_{\{k_i\}}\right)\otimes e^{(t-\tau_1^+)\mathcal L_2}\left(\rho_{\{k_i\}}(\tau_1^+)\right),
\end{split}
\end{equation}
so that using eq.~\eqref{eq:state tau1+} we obtain the desired result 
\begin{equation}
\Tr_T\left[e^{t\mathcal L}(\phi_0\otimes\rho_0)\right]\simeq e^{(t-\tau_1^+)\mathcal L_2}e^{\tau_1^- \mathcal L_1}(\rho_0) + \err{\eps N}.
\end{equation}
To obtain eq.~\eqref{eq:tau1+<t<tau} we need to show that $[\mathcal L_T,\mathcal L_S]=0$ when applied to $\phi_{\{k_i\}}\otimes\rho$.
Indeed, defining the projector $P_{T_1,T} = \sum_{k_1,\dots,k_N = T_1}^{T-1}\phi_{\{k_i\}}$  and its relative superoperators 
$\mathcal P_{T_1,T}(A) = P_{T_1,T}\,A\, P_{T_1,T}$, a direct computation shows that $\mathcal P_{T_1,T}[\mathcal L_T,\mathcal L_S]\mathcal P_{T_1,T}=0$.
Then, for $\tau_1^+ \leq t \leq \tau^-$,
\begin{equation}
\label{eq:tau1+<t<tau projector}
\begin{split}
\mathcal P_{T_1,T}\, e^{t\mathcal L}&(\phi_0\otimes\rho_0)
\simeq \sum_{k_1,\dots, k_N = T_1}^{T-1} p_{\{k_i\}}(\tau_1^+) \mathcal P_{T_1,T}\,e^{(t-\tau_1^+)\mathcal L}\mathcal P_{T_1,T} \left(\phi_{\{k_i\}}\otimes\rho_{\{k_i\}}(\tau_1^+)\right) \\
&\simeq \sum_{k_1,\dots, k_N = T_1}^{T-1} p_{\{k_i\}}(\tau_1^+)\,\left(\mathcal P_{T_1,T}\,e^{(t-\tau_1^+)\mathcal L_T}\left(\phi_{\{k_i\}}\right)\right)\otimes e^{(t-\tau_1^+)\mathcal L_2}\left(\rho_{\{k_i\}}(\tau_1^+)\right).
\end{split}
\end{equation}
Moreover, we can remove the projector on the lhs by noticing from eq.~\eqref{eq:tau1+<t<tau initial} that $\mathcal P_{T_1,T}\, e^{t\mathcal L}(\phi_0\otimes\rho_0)\simeq e^{t\mathcal L}(\phi_0\otimes\rho_0)$.
We would like to remove $\mathcal P_{T_1,T}$ also from the rhs.
To do this, let us compute the evolution of the timers by tracing out the system,
\begin{equation}
\label{eq:tau1+<t<tau time evolution}
\begin{split}
\Tr_S\left[ \mathcal P_{T_1,T}\, e^{t\mathcal L}(\phi_0\otimes\rho_0)\right]
&\simeq \mathcal P_{T_1,T}\, e^{(t-\tau_1^+)\mathcal L_T} \sum_{k_1,\dots, k_N = T_1}^{T-1} p_{\{k_i\}}(\tau_1^+)\,\phi_{\{k_i\}} \\
&\simeq \mathcal P_{T_1,T}\, e^{(t-\tau_1^+)\mathcal L_T} e^{\tau_1^+\mathcal L_T}(\phi_0) \simeq \mathcal P_{T_1,T}\,e^{t\mathcal L_T}(\phi_0).
\end{split}
\end{equation}
For $\tau_1^+ \leq t \leq \tau^-$, eq.~\eqref{eq:tau1+<t<tau prob} tells us that we can remove the projector in the last expression by simply picking up error terms exponentially small in $M$.
This last result allows us to remove $\mathcal P_{T_1,T}$ in the last line of eq.~\eqref{eq:tau1+<t<tau projector} and finally get eq.~\eqref{eq:tau1+<t<tau}.
In particular, specializing eq.~\eqref{eq:tau1+<t<tau} to $t=\tau^-$ and tracing out the timers
\begin{equation}
\label{eq:tau1+<t<tau traceT}
\begin{split}
\Tr_T\left[e^{\tau^-\mathcal L}(\phi_0\otimes\rho_0)\right] 
&\simeq e^{(\tau^- -\tau_1^+)\mathcal L_2}\left(\sum_{k_1,\dots, k_N = T_1}^{T-1} p_{\{k_i\}}(\tau_1^+)\,\rho_{\{k_i\}}(\tau_1^+)\right) \\
&\simeq e^{(\tau^- -\tau_1^+)\mathcal L_2} e^{\tau_1^-\mathcal L_1}(\rho_0) + \err{\eps M},
\end{split}
\end{equation}
where in the last line we used eq.~\eqref{eq:state tau1+}.

The second $\eps$-time window $\tau^\pm$ is treated as the previous one,
\begin{equation}
\label{eq:window tau}
\begin{split}
\Tr_T\left[e^{\tau^+ \mathcal L}(\phi_0\otimes\rho_0)\right] 
&\simeq \Tr_T\left[e^{\tau^- \mathcal L}(\phi_0\otimes\rho_0)\right] + \err{\eps M} \\
&\simeq e^{(\tau^- -\tau_1^+)\mathcal L_2} e^{\tau_1^-\mathcal L_1}(\rho_0) + \err{\eps M} \\
&\simeq e^{(\tau -\tau_1)\mathcal L_2} e^{\tau_1\mathcal L_1}(\rho_0) + \err{\eps M},
\end{split}
\end{equation}
where in the second line we used eq.~\eqref{eq:tau1+<t<tau traceT}, and in the last line we substituted $\tau_1^\pm$ with $\tau_1$ and $\tau^-$ with $\tau$ without changing the order of the error.

Finally, consider late times $t\geq \tau^+$.
The state can be written as in eq.~\eqref{eq:transitivity after tau},
\begin{equation}
\label{eq:after tau}
e^{t\mathcal L}(\phi_0\otimes\rho_0) = 
\sum_{\{k_i\}_i\neq \{T\}_i} p_{\{k_i\}}(t)\, \phi_{\{k_i\}} \otimes \rho_{\{k_i\}}'' + p_T(t)\, \phi_T\otimes \rho'(t).
\end{equation}
Now the timers are all almost surely in the state with all switches flipped to one, therefore the first term of eq.~\eqref{eq:after tau} is exponentially small in the thermodynamic limit and we are left with the second one, $e^{t\mathcal L}(\phi_0\otimes\rho_0) \simeq  \phi_T\otimes \rho'(t)$.
To find an expression for $\rho'(\tau^+)$ we trace out the timers and use the previous result eq.~\eqref{eq:window tau},
\begin{equation}
\rho'(\tau^+) = \Tr_T\left[ e^{\tau^+\mathcal L}\left(\phi_0\otimes\rho_0\right) \right]
\simeq e^{(\tau^- -\tau_1)\mathcal L_2} e^{\tau_1\mathcal L_1}(\rho_0) + \err{\eps M}.
\end{equation}
Finally, after $\tau^+$ the whole system is not evolving any more,
\begin{equation}
\begin{split}
e^{t\mathcal L}(\phi_0\otimes\rho_0) 
&\simeq e^{(t-\tau^+)\mathcal L} \left(\phi_T\otimes e^{(\tau -\tau_1)\mathcal L_2} e^{\tau_1\mathcal L_1}(\rho_0)\right)  + \err{\eps M} \\
&\simeq \phi_T\otimes e^{(\tau -\tau_1)\mathcal L_2} e^{\tau_1\mathcal L_1}(\rho_0) + \err{\eps M} .
\end{split}
\end{equation}
The ancillas are in the fixed point of their evolution and are in a product state, so they can be easily discarded.
The final state of the system is the desired one, and the error goes to zero in the thermodynamic limit.

\section{The other implication: Comments and examples}
\label{sec:other implication}

\subsection{Unique SPT phase in one dimension: a difference with the Hamiltonian classification}
\label{subsec:1D SPT phase}

In this section we will show that our definition of phases differs from the ones through gapped paths of Hamiltonian, at least in one dimension and when we consider only states with a specific symmetry described by a group $G$, and evolutions that are covariant under the symmetry.

Symmetry Protected Topological (SPT) phases in one dimension can be studied with Matrix Product States (MPS), as done in Ref.~\cite{Schuch2011,Garre-Rubio2017}.
This is possible since any ground states of gapped quantum Hamiltonians can be approximated efficiently with MPS.
In one dimension, if no symmetry restriction is imposed, all MPS turn out to be in the same phase.
However, if we require the states and the Hamiltonian path to be invariant under certain on-site global linear symmetries, then the phase diagram becomes richer.
In the case of injective MPS, when their parent Hamiltonian has a single ground state, the different phases are labelled by the second cohomology group $H^2(G,U(1))$ of the symmetry group $G$ over $U(1)$.
To understand why, consider the action of the on-site symmetry on the MPS projector $\mathcal P$.
The global symmetry acts as a unitary representation of the symmetry group $U_g$, $g\in G$, acting on the physical index of $\mathcal P$.
It is well known~\cite{perez2007matrix} that it is possible to bring any MPS to a standard form where the action of the symmetry on the physical index translates to an action of a projective unitary representation of $G$ on the virtual indices, namely
\begin{equation}
\label{eq:P symm}
U_g\mathcal{P} = \mathcal{P}(V_g\otimes\bar V_g),
\end{equation}
where the bar denotes the complex conjugate.
The $V_g$ do not need to be a linear representation of $G$, since they always appear together with their complex conjugate, and therefore any extra phase would cancel.
For this reason the $V_g$ are determined up to a phase $V_g\sim e^{\iu\chi_g} V_g$.
Since the $V_g$ are projective representations $V_gV_h = e^{\iu\omega(g,h)}V_{gh}$, and for the same reason also $\omega(g,h)$ is defined up to the equivalence relation
\begin{equation}
\label{eq:equivalence relation}
\omega(g,h)\sim\omega(g,h)+\chi_{gh}-\chi_g-\chi_h.
\end{equation}
The equivalence classes defined by the previous relations form a group isomorphic to $H^2(G,U(1))$.
In Ref.~\cite{Schuch2011} it is shown that two MPS with the same cohomology class for $V_g$ can be connected by a path of gapped Hamiltonian respecting the symmetry $G$, that is such that $[U_g^{\otimes N},H(s)]=0$, for $s\in[0,1]$.
On the contrary, for two MPS with different cohomology classes, such path cannot exist.

For our purposes, it is enough to consider the so called isometric form of the MPS, some specific representatives of any phases of MPS, that are fixed point of certain renormalization procedures~\cite{Schuch2010}.
For the case of a single non-degenerate ground state, the isometric form is just a tensor product of maximally entangled pairs of dimension $D$ (the bond dimension) shared by neighbouring sites, $\ket{\phi_D}=\ket*{\phi_{+}^{D}}^{\otimes N}$.
On site $i$ the physical Hilbert space $\mathcal H_i$ is the tensor product of two Hilbert spaces of dimension $D$, the left Hilbert space $\mathcal H_{i,L}$ and the right Hilbert space $\mathcal H_{i,R}$.
The $i$-th maximally entangled state $\ket*{\phi_{+}^{D}}_{i}$ lives in $\mathcal H_{i,R}\otimes \mathcal H_{i+1,L}$.
The symmetry $G$ is realized on $\ket{\phi_D}$ as $U_g^{\otimes N}$ with the on-site symmetry realization $U_g= V_g\otimes\bar V_g$, for any projective representations $V_g$.
	Notice that $U_g$ is a linear representation, not a projective one, and that eq.~\eqref{eq:P symm} is evidently satisfied since in this case $\mathcal P=\mathbb I$.
Since projective representation are defined up to the equivalence relation of eq.~\eqref{eq:equivalence relation}, we can label the different isometric MPS with the elements of $H^2(G,U(1))$.
Let us call these states $\ket{\phi_D,\omega}$, for $\omega \in H^2(G,U(1))$.
In Ref.~\cite{Schuch2011} it is shown that any state that belongs to the cohomology class $\omega$ can be connected by a smooth path of gapped Hamiltonians to $\ket{\phi_D,\omega}$.
When taking the $\ket{\phi_D,\omega}$'s as representatives of the different phases, we are actually fixing the state $\phi_D$, and changing the realization of the symmetry group $G$.
Notice that in this framework it is possible to compare states that transform with different representations of the symmetry group.
In this case one can take as Hilbert space the direct sum of the Hilbert spaces of the two systems (plus some environment space that may be accessed during the path) $\mathcal H = \mathcal H_{0} \oplus \mathcal H_{1}\oplus \mathcal H_\text{path}$.
The initial Hamiltonian $H_0$ is supported on $\mathcal H_{0}$ while $H_1$ is supported on $\mathcal H_{1}$.
The Hamiltonian along the path is invariant under a representation $U_g=U_g^0\oplus U_g^1\oplus U_g^\text{path}$.
This can be achieved by adding some local ancillas to the system, as commented in Section~\ref{sec:pure case}, after Definition~\ref{def:H definition}.

Here we show that from the dissipative point of view SPT phases are not protected by symmetry any more.
In other words, any state can be driven fast to any other in a different SPT phase with a dissipative evolution which is covariant under the action of the realization of the symmetry group $U_g$, namely 
\begin{equation}
\label{eq:L covariant}
U_g^\dag\, \mathcal L(U_g \,\cdot\, U_g^\dag)\,U_g = \mathcal L(\,\cdot\,).
\end{equation}
In particular $\mathcal L$ sends invariant states to invariant states.
Notice that the previous requirement is implied for example in case the Hamiltonian $H$ and the Lindblad operators $L_k$ are invariant under $U_g$, namely $U_g^\dag H U_g = H$ and $U_g^\dag L_k U_g = L_k$.
To prove the claim we need two results:
First, we need to show that any two states that are in the same cohomolgy class, and therefore in the same Hamiltonian phase, are still in the same phase according to the Lindbladian definition.
We will consider the fast Lindbladian constructed from the Hamiltonian path in Section~\ref{sec:pure case}, and check that it has the required transformation properties under the action of the symmetry group.
Second, we show there exists a fast Lindbladian evolution, invariant under any representations of $G$ of the form $V_g\otimes \bar V_g$, such that it connects two specific representatives (obtained by tensor products of the isometric MPS) of the phases labelled by $\omega,\omega'\in H^2(G,U(1))$.

Let us start from the latter task, by finding a fast Lindbladian that connects two representatives of phases with different cohomology class.
Before proceeding, we need an intermediate result.
We need to show that it is possible to drive fast any state to a product state $\ketbra{\psi}^{\otimes N}$.
This is simply done by defining a single site CPTP map $\mathcal T_i (\cdot) = \Tr_i[\,\cdot\,] \ketbrasub{\psi}{\psi}{i}$, where $\Tr_i[\,\cdot\,]$ is the partial trace over site $i$.
From such a CPTP map we can build the single site Lindbladian $\mathcal L_i = \mathcal T_i - \id$ and finally $\mathcal L = \sum_i \mathcal L_i$.
The action on any state of the dissipative evolution arising from a single site term can be easily computed after observing that $\mathcal T_i^2 = \mathcal T_i$.
One finds $e^{t(\mathcal T_i - \id)}(\rho) = \left(\mathcal T_i - e^{-t}\mathcal L_i\right)(\rho)$.
The evolution on different sites trivially commutes, therefore
\begin{equation}
\label{eq:fast local L}
e^{t\mathcal L}(\rho)=e^{t\sum_i(\mathcal T_i-\id)}(\rho) = \prod_{i\in\Lambda}\left[(1-e^{-t})\mathcal T_i + e^{-t}\right] (\rho) =
\sum_{S\subseteq\Lambda}e^{-|\compl{S}|t}(1-e^{-t})^{|S|}\mathcal T_S(\rho),
\end{equation}
where $\mathcal T_S(\rho)= \big[\prod_{i\in S}\mathcal T_i\big](\rho)$.
From eq.~\eqref{eq:fast local L} it is clear that the only term in the sum that does not vanish in the limit of large $t$ is the one for which $S=\Lambda$.
Then, $\lim_{t\to\infty} e^{t\mathcal L}(\rho) = \mathcal T_\Lambda (\rho)$.
Moreover, by direct inspection of eq.~\eqref{eq:fast local L} one can see that for positive times, $\norm*{e^{t\mathcal L}(\rho)-\mathcal T_\Lambda (\rho)}_1\lesssim 1-\left(1-e^{-t}\right)^N$.
This last quantity vanishes in the thermodynamic limit if $t\gtrsim\poly\log N$ for any initial state, and the Lindbladian is therefore fast mixing%
\footnote{Notice that this Lindbladian has a single pure stationary state, therefore we could have proven fast mixing by directly applying Proposition 10 of Ref.~\cite{Kastoryano2012}.}.
This analysis is actually valid for any tensor product of single site Lindbladians obtained from a single site idempotent CPTP map.
In our particular case, $\mathcal T_\Lambda(\rho) = \ketbra{\psi}^{\otimes N}$ and this proves our claim.

In our setting, we are interested in the case in which $\ket{\psi}_i=\ket{\psi_+}_{i}$ lives on site $i$ and is the maximally entangled state between $\mathcal H_{i,L}$ and $\mathcal H_{i,R}$.
The state $\ket{\psi_+}^{\otimes N}$ clearly belongs to the trivial phase, since it is invariant for any unitary of the form $U\otimes\bar U$ acting on any site $i$, and in particular the realization of the symmetry $U_g=V_g\otimes\bar V_g$ acts trivially $U_g\ket{\psi_+} = \ket{\psi_+}$.
Notice also that for the same reason the single site Lindbladian $\mathcal L_i(\cdot) = \Tr_i[\,\cdot\,] \ketbrasub{\psi_+}{\psi_+}{i} - \id(\cdot)$ is covariant according to eq.~\eqref{eq:L covariant}.
One also has to use the cyclicity of the trace.
For $U_g$ acting on site $i$ we have $\Tr_i[U_g\rho U^\dag_g]\,U^\dag_g \ketbrasub{\psi_+}{\psi_+}{i} U_g = \Tr_i[U^\dag_gU_g\rho]\ketbrasub{\psi_+}{\psi_+}{i} = \ketbrasub{\psi_+}{\psi_+}{i}$, while when the symmetry acts on $j\neq i$ the covariance is trivially checked.

To conclude our argument consider the two following states obtained by tensoring three states of the kind described so far
\begin{subequations}
\label{eq:phi_0 and phi_1}
\begin{align}
\ket{\phi_0} &= \ket{\phi_D,\omega_0} \otimes \ket*{\phi_D,\omega_1^{-1}} \otimes \ket{\phi_D,\omega_1}, \\
\ket{\phi_1} &= \ket{\psi_+}^{\otimes N} \otimes \ket{\psi_+}^{\otimes N} \otimes \ket{\phi_D,\omega_1}.
\end{align}
\end{subequations}
Here $\ket*{\phi_D,\omega_1^{-1}}$ is the isometric MPS representative of the phase labelled by the inverse of the element of the second cohomology group $\omega_1$.
Therefore, the state $\ket{\phi_0}$ belongs to the phase labelled by $\omega_0$.
It can be shown that there exists a path of gapped Hamiltonian connecting $\ket{\phi_0}$ with $\ket{\phi_D,\omega_0}$.
On the other hand, the state $\ket{\phi_1}$ is a tensor product of the representative of the phase labelled by $\omega_1$ times two tensor product states which belong to the trivial phase.
Therefore it belong to the phase labelled by $\omega_1$.
Finally, we prove our claim by applying the evolution governed by the covariant Lindbladian $\mathcal L$ to the first and second factors in $\ket{\phi_0}$, while keeping the third factor untouched.
The Lindbladian will evolve in short times the state $\ket{\phi_0}$ to the state $\ket{\phi_1}$, thus proving that we can connect to each other two representatives of any (Hamiltonian) SPT phases.

What is left to show is that any two states in the same SPT phase according to the Hamiltonian definition, are still in the same phase according to the Lindbladian definition.
In order to do so, let us consider the Lindbladian built in Section~\ref{sec:pure case} from the gapped path of Hamiltonian.
By hypothesis, for any two states in the same SPT phase such gapped path does exist.
In Section~\ref{sec:pure case} we considered the related quasi-adiabatic evolution, we approximated it with a unitary quantum circuit of finite depth, and finally we built a Lindbladian by adding some ancillary systems acting as timers and triggering at specific times the action of the gates of the circuit.

We now briefly go through these steps to check that the resulting Lindbladian is covariant according to eq.~\eqref{eq:L covariant}.
We need here an important remark, namely that we do not require the ancillary system to be invariant under the group symmetry $G$.
So, only the action of $\mathcal L$ on the original system has to be covariant under the action of $G$.
First of all, from eq.~\eqref{eq:quasiadiabatic K} we see that the quasi-adiabatic Hamiltonian $K(s)$ is built from $H(s)$ and therefore it is evidently invariant under the representation $U_g$.
Also the terms $k_{j,\alpha}(s)$ as defined in eq.~\eqref{eq:kjalpha def} are invariant.
Indeed they are defined through linear combinations of terms that involve the local terms of the Hamiltonian $h_{j,\alpha}(s)$, which are all independently invariant.
The gates of the first layer of the quantum circuit are of the form $e^{i\int_0^1 K_A(s)}$ with $K_A(s)$ defined in eq.~\eqref{eq:K_A def} as the sum of the terms $k_{j,\alpha}(s)$ completely supported on $A$, and therefore they are invariant.
For the second layer, starting from the $k_{j,\alpha}(s)$ we have built the $\ell_{j,\alpha}(s)$ defined in eq.~\eqref{eq:ell def}, as sum of terms which are all invariant under the symmetry action $U_g$.
See also the definition in eq.~\eqref{eq:L Delta} and below.
According to the dimensionality of the system and depending on how we partition the lattice in regions that are the supports of the gates, as already discussed at the end of Section~\ref{subsec:finite depth circuit}, we may need to split the gates of these two layers in smaller ones.
For these successive layers, we need to build some new $\ell'_{j,\alpha}(s)$ from the $\ell_{j,\alpha}(s)$ in the same way as the $\ell_{j,\alpha}(s)$ are defined through the $k_{j,\alpha}(s)$, and continue this procedure as many times as needed to finally have gates that are supported on sets of size $\sim\poly\log N$.
It is clear however that this procedure will generate gates that are invariant under $U_g$. 

As a last step, we simply notice that the action of the Lindbladian $\mathcal L$ defined at the beginning of Section~\ref{subsec:L from circuit} on the original system is through a Hamiltonian evolution $h_i^{(\ell)}$ relative to the unitary gate $U^{(\ell)}_{i}$ that belong to one of the layers.
Since $U^{(\ell)}_{i}$ is invariant under the action of $U_g$, so we can choose $h_i^{(\ell)}$.
When applying $\mathcal L$ to any state $\rho$, the actual set of $M$ gates of the circuit that is applied depends on the internal state of the timers.
However, for any state of the timer, the action is covariant and this concludes the proof.

\subsection{An example of two-dimensional topological states that are not in the same phase}
\label{subsec:slow Zm to Zn}

In this section we will show that, at least in some particular cases, there is no Lindbladian that goes fast (according to eq.~\eqref{eq:formal definition}) from a pure state with a certain topological order to a pure state whose topological order contains the one of the previous state.
Specifically, we will consider two quantum double models, $D(\mathbb{Z}_n)$ and $D(\mathbb{Z}_m)$ with $m<n$.
In particular, the result will be valid for $m$ a divisor of $n$.
In this case $\mathbb Z_m$ is a (normal) subgroup of $\mathbb Z_n$, and in this sense the $\mathbb Z_m$-topological order is contained in the $D(\mathbb Z_n)$ model.	
We will follow the line of reasoning of Ref.~\cite{Konig2014}.
There, the authors show that with a local Lindbladian evolution it is not possible to generate certain topological ordered states from a product state faster than in a time proportional to the linear system size.
We extend their results allowing for some topological order in the initial state.
However, while the results of Ref.~\cite{Konig2014} are valid for any (pure or mixed) initial and final states, the results presented here are valid only in the case of pure states.
Moreover, as for~\cite{Konig2014} the results are limited to the case of geometries which exhibit ground state degeneracy.
We will later explain which are the technical problems in dealing with the more general case.

\begin{main}
Given two pure states $\ket{\phi}\in\GS{m}$ and $\ket{\psi}\in\GS{n}$, there is no local Lindbladian $\mathcal L$ as defined in Definition~\ref{def:L definition} such that $\phi\fastto[\mathcal L] \psi$ if $m<n$.
Any such dissipative evolution from $\phi$ to $\psi$ would require a time at least of the same order of the system size.
\end{main}

\paragraph{Setup.}
Consider $\psi=\ketbra{\psi}$, a pure state supported on $\GS{n}$, the ground state space of the $D(\mathbb{Z}_n)$ model.
For concreteness, we take the double model on a torus with linear size $L$ in both directions and with $N\propto L^2$ sites.
In this geometry $\GS{n}$ has dimension $n^2$.
As it will be clear from the discussion, the results can be easily generalized to geometries with any genus, planar geometries with boundaries, and in general to any geometries with ground state degeneracy.
Let us then introduce the unitary logical operator $\lX=\prod_{i\in p^\star}X_i$ that is supported on a path $p^\star$ in the dual lattice along a non trivial loop of the torus.
The explicit path is not important, since this operator is equivalent (in the ground state space algebra) to any other supported on any loops in the same homotopy class.
Therefore, in a geometry with genus $g$ there are $2^g$ independent $\lX$ operators.
For concreteness, on the torus let us take two representatives $\lX_x$ and $\lX_y$ supported respectively on a vertical and a horizontal straight line at position $x$ and $y$.
Its powers $\lX^\alpha$, for $\alpha=0,\dots,n-1$ form a representation of the group $\mathbb{Z}_n$.
To generate the full ground state space algebra we also need the logical unitary operators $\lZ=\prod_{i\in p}Z_i$, that are supported on loops $p$ in the original lattice.
Again we choose two representatives $\lZ_x$ and $\lZ_y$ on the torus\footnote{Strictly speaking the $x$ and $y$ coordinate here are not the same as the ones of $\lX_x$ and $\lX_y$, since the former are coordinates on the dual lattice, and the latter are coordinates of the real lattice.
However, since in the following this distinction is never used, we will use the same symbols for simplicity.}.
These operators form another representation of $\mathbb{Z}_n$ and their commutation relations with the $\lX_{x,y}$ are given by the relations $\lX_x \lZ_y = \omega_n \lZ_y \lX_x$ and $\lX_y \lZ_x = \omega_n \lZ_x \lX_y$, with $\omega_n=e^{\frac{2\pi\iu}{n}}$.
Consider now a basis $\ket{\alpha,\beta}$ of eigenvectors of $\lZ_x$ and $\lZ_y$, such that $\lZ_x^\gamma\ket{\alpha,\beta}=\omega_n^{\beta\gamma}\ket{\alpha,\beta}$ and $\lZ_y^\delta\ket{\alpha,\beta}=\omega_n^{\alpha\delta}\ket{\alpha,\beta}$.
On this basis the $\lX_{x,y}$ act as ladder operators, $\lX_x^\gamma\ket{\alpha,\beta}=\ket{\alpha+\gamma,\beta}$ and $\lX_y^\delta\ket{\alpha,\beta}=\ket{\alpha,\beta+\delta}$.
These operators generate the full $n^{4g}$-dimensional algebra of $\GS{n}$.

We will prove the main result of this section by contradiction.
To this purpose, for any two pure states $\psi=\ketbra{\psi}$ supported on $\GS{n}$ and $\phi=\ketbra{\phi}$ supported on $\GS{m}$ with $m<n$, we suppose there exists a Lindbladian $\mathcal L$ such that $\phi\fastto[\mathcal L]\psi$.

\paragraph{Generating the full ground state space from $\ket{\psi}$.}
Before proceeding with the proof, we need an intermediate result which essentially allows us to work with only $n^{2g}$ operators, in the case at hand either the $\lX_{x,y}^\alpha$ for $\alpha=0,\dots,n-1$, or the $\lZ_{x,y}^\alpha$'s.

\begin{step}
Given any state in the ground state space of the $D(\mathbb Z_n)$ model $\ket{\psi}\in \GS{n}$, at least one of the two sets $\psi_X=\{\ket{\psi_i}=\ldX^i\ket{\psi}\}_{i=0}^{n-1}$ and $\psi_Z=\{\ket*{\psi^\prime_j}=\ldZ^j\ket{\psi}\}_{j=0}^{n^2-1}$ is a basis of $\GS{n}$.
\end{step}

First of all, we notice that we can build some $\ldX$ and $\ldZ$ operators on the full $n^2$ dimensional space using respectively the $\lX_{x,y}$ and the $\lZ_{x,y}$.
We order the vectors of the eigenbasis of the $\lZ_{x,y}$'s as $\ket{\alpha,\beta}\cong\ket{\alpha+n\,\beta}$, and defining $\ldX^i=\lX_x^\gamma \lX_y^\delta$ and $\ldZ^i=\lZ_x^\delta \lZ_y^{\delta+\gamma/n}$, with $i=\gamma+n\,\delta$, we can easily check that they have all the required properties.
The $\ldZ^i$ are diagonal on the described basis $\ldZ^i \ket{j} = \omega_{n^2}^{ij}\ket{j}$, and the $\ldX^i$ act as ladder operator $\ldX^i \ket{j} = \ket{j+i}$.
The second step consists in showing that starting from a vector $\ket{\psi}$ in $\GS{n}$, it is possible to build a set of $n^2$ independent vectors by applying either the $\ldX^i$ or the $\ldZ^i$.
Consider the two sets of states $\psi_X=\{\ket{\psi_i}=\ldX^i\ket{\psi}\}_{i=0}^{n^2-1}$ and $\psi_Z=\{\ket*{\psi^\prime_j}=\ldZ^j\ket{\psi}\}_{j=0}^{n^2-1}$.
We show now that at least one of them is composed by linearly independent vectors.
We can prove this claim by contradiction:
suppose that in both sets the vectors are linearly dependent.
Then the vector $\psi$ can be written both as 
$\ket\psi = \sum_{i=1}^{n^2-1}a_i\ket{\psi_i}=\big(\sum_{i=1}^{n^2-1}a_i\ldX^i\big)\ket{\psi}$ and 
$\ket\psi = \sum_{j=1}^{n^2-1}b_j\ket*{\psi^\prime_j}=\big(\sum_{j=1}^{n^2-1}b_j\ldZ^j\big)\ket{\psi}$.
Therefore $\ket{\psi}$ is an eigenvector of both the operators $A=\sum_{i=1}^{n^2-1}a_i\ldX^i$ and $B=\sum_{j=1}^{n^2-1}b_j\ldZ^j$.
First of all, notice that these two operators have the same eigenvectors of respectively $\ldX$ and $\ldZ$.
If neither the spectrum of $A$ nor the one of $B$ have degeneracies, since $\ldX$ and $\ldZ$ have no common eigenvectors, then we already have a contradiction.
The only possibility left is the case when one of the two, say for example $B$, has a degenerate eigenspace of dimension $d>1$, and $\ket{\psi}$ belongs to such a subspace.
Since the eigenvectors of $B$ are the eigenvectors of $\ldZ$, then $\ket{\psi}$ belongs to a subspace spanned by $d$ eigenvectors of $\ldZ$, namely $\ket{\psi}=\sum_{j\in D}c_j\ket{j}$, with $\ldZ\ket{j}=\omega^j\ket{j}$ and $D=\{j : B\ket{j}=\ket{j} \}$.
Notice that some of the $c_j$ with $j\in D$ may vanish, but at least two of them should be different from zero (if only one is different from zero, then $\ket{\psi}$ is an eigenvector of $\ldZ$ and therefore the $\ket{\psi_i}$'s are an orthonormal basis).
Also, the $d$-dimensional invariant subspace cannot be the full $n$-dimensional space, because in that case one would have $B=\mathbb{I}$, which is impossible since $\Tr[B]=0$.
Therefore we have $1<d<n$.
Notice that the vector $\ket{\psi}$ belonging to a (proper) subspace invariant under $\ldZ$ is indeed equivalent to the $\ket{\psi'_{i}}$'s being linearly dependent.
Let us now apply the hypothesis that $\ket{\psi}$ is an eigenvector of $A$.
Since the operators $\ldX^i$ act on the $\ket{j}$ as ladder operators, the eigenvector relation reads $\sum_{j\in D}\sum_{i=1}^{n^2}c_j\, a_i \ket{j+i}=\sum_{j\in D}c_j\ket{j}$.
By taking the inner product with $\ket{k}$ we can check it explicitly component by component, $c_k\,\delta_{k\in D} = \sum_{j\in D}a_{k-j}\,c_j,\,\forall k=0,\dots,n^2-1$.
To find a contradiction, we want to show that this set of equations for the $a_i$'s and the $c_j$'s has no solution.
We can rewrite the problem has an eigenvalue problem in $\mathbb{R}^n$, $v=M\cdot v$.
The vector $v$ has components $v_i=c_i$ for $i\in D$ and zero elsewhere.
Our problem only fixes the columns of the matrix $M$ whose indexes belong to $D$, namely $M_{i,j}=a_{i-j}$ if $j\in D$ and where we defined $a_0=0$.
We can complete the matrix $M$ with anything in the remaining columns, since they will always be multiplied by the zero components of $v$.
Any solution of the original problem is a solution of the `completed' problem.
In particular, we can complete $M$ such that it is circulant, by setting $M_{i,j}=a_{i-j},\,\forall i,j=0,\dots,n^2-1$.
We are left with the problem of finding a circulant matrix with at least an eigenvector with $d$ components equal to zero.
This is the contradiction, since all eigenvectors $w^{(j)}$, $j=0,\dots,n^2-1$, of circulant matrices have components $w^{(j)}_i \propto \omega_{n^2}^{ij},\,i=0,\dots,n^2-1$.

We have seen that at least one between $\psi_X$ and $\psi_Z$ is a set of linearly independent vectors.
Without loss of generality, consider this to be $\psi_X$, and consider the overlap matrix $S=\{\braket{\psi_{\alpha,\beta}}{\psi_{\gamma,\delta}}\}_{\alpha,\beta,\gamma,\delta=0}^{n-1}$, with $\ket{\psi_{\alpha,\beta}}=\lX_x^\alpha \lX_y^\beta\ket{\psi}$.
Since the vectors are linearly independent, such matrix (with $(\alpha,\beta)$ as row index and $(\gamma,\delta)$ as column index) has full rank.
Using the fact that $\mathcal L$ is (quasi,$\log$)-local and drives fast $\phi$ to $\psi$, we can build a similar full rank matrix of vectors in $\GS{m}$, thus providing a contradiction since $\GS{m}$ has dimension $m<n$.

\paragraph{Dissipative evolution of the ground state space.}
We need another intermediate result, about the evolution under $\mathcal L$ of any state of $\GS{m}$ other than $\phi_0$.
It turns out that after a time which scales $\poly$-logarithmic in the system size the evolved states are close to states supported in $\GS{n}$.

\begin{step}
Consider a time $t\gtrsim \poly\log N$ such that $\norm*{e^{t\mathcal L}(\phi)-\psi}_1\leq \eps_N$, for vanishing $\eps_N$ in the limit $N\to\infty$.
Under our hypothesis the  evolution under $\mathcal L$ of any pure state $\ket*{\tilde{\phi}}\in\GS{m}$ after the same time $t$ is close to 
a state $\tilde{\psi}$ supported in $\GS{n}$, namely $\norm*{e^{t\mathcal L}(\tilde\phi)-\tilde\psi}_1\leq \eps_N'$, for some vanishing $\eps_N'$ in the thermodynamic limit.
\end{step}

To prove this, we just need to show that $\Trace[e^{t\mathcal L}(\tilde\phi)\, H^{(n)}]=O(\eps_N)$, with $H^{(n)}$ any local frustration free Hamiltonian whose ground state space is $\GS{n}$.
Since $H^{(n)}$ is local we can write it as a sum $H^{(n)}=\sum_i h_i^{(n)}$ of terms with finite support, $\supp\left[h_i^{(n)}\right]=S_i$, $|S_i|<k$ with $k$ constant in $N$.
Then, by duality we can consider the evolved of the $h_i^{(n)}$'s and use the locality of $\mathcal L$ and Lieb-Robinson bounds to approximate them with operators whose support is of size $\poly\log N$.
More precisely,
\begin{equation}
\label{eq:GSS evolution}
\Tr[h_i^{(n)}\,e^{t\mathcal L}(\tilde{\phi})]=\Tr[e^{t\mathcal L^*}(h_i^{(n)})\,\tilde\phi] = \Tr[\fat[\ell]{h_i^{(n)}}\,\tilde\phi]+\err{kG(\ell)},
\end{equation}
where $\fat[\ell]{A}$ is the `fattening' of the operator $A$, namely the evolution of $A$ according to the Lindbladian obtained retaining only the terms of $\mathcal L^*$ that are fully supported in the $\ell$-ball of the support of $A$.
In formulas, $\fat[\ell]{A}=e^{t\mathcal L^*_{S_\ell}}(A)$, with $S_\ell(A)$ the `fattening' of the support of $A$, $S_\ell(A) = \{x:d(x,\supp[A])\leq\ell\}$.
The error in the last expression is the one given by Lieb-Robinson bounds, with the function $G(\ell)$ that vanishes for large $\ell$.
Its explicit form is not necessary and as seen in the previous sections, in the case considered in this work if we choose $\ell\sim\poly\log L$ it will decay faster than any polynomial in $L$ and the error will vanish in the thermodynamic limit $L\to\infty$.
Notice now that the initial pure state $\ket*{\tilde \phi}$ can be obtained from $\ket{\phi}$ by applying logical operators, $\ket*{\tilde \phi}=L_xL_y\ket{\phi}$.
In general these may be combinations of the $\lX_{x,y}^{(m)}$'s and the $\lZ_{x,y}^{(m)}$'s generating the logical algebra on $\GS{m}$, but in any case they are supported on straight lines along the fundamental cycles of the torus.
Since the exact positions $x$ and $y$ are immaterial and can be chosen at will, let us pick them such that $d(\supp[L_x],\supp[h_i^{(n)}]),\,d(\supp[L_y],\supp[h_i^{(n)}])>\ell$.
With this choice the logical operators trivially commute with the fattening of the Hamiltonian term $h_i^{(n)}$, hence using the cyclicity of the trace $\Tr[\fat[\ell]{h_i^{(n)}}\,\tilde\phi]=\Tr[\fat[\ell]{h_i^{(n)}}\,\phi]$.
By reversing the passages in eq.~\eqref{eq:GSS evolution},
\begin{equation}
\label{eq:GSS evolution 2}
\Tr[\fat[\ell]{h_i^{(n)}}\,\phi] = \Tr[e^{t\mathcal L^*}(h_i^{(n)})\,\phi]+\err{kG(\ell)} = \Tr[h_i^{(n)}\,e^{t\mathcal L}(\phi)]+\err{kG(\ell)}.
\end{equation}
By following the same procedure for any Hamiltonian term $h_i^{(n)}$, and using the hypothesis on the evolution of $\phi$, we conclude that
\begin{equation}
\label{eq:GSS evolution 3}
\Tr[H^{(n)}\,e^{t\mathcal L}(\tilde{\phi})] = \Tr[H^{(n)}\,\psi]+\poly(N)\cdot\err{kG(\ell),\eps_N}.
\end{equation}
The factor $\poly(N)$ in the right hand side comes from the number of terms in the Hamiltonian.
The error vanishes in the thermodynamic limit, thus proving what we needed.

We now make two comments.
First of all, this result can be immediately extended to the case in which $\tilde{\phi}$ is a mixed state supported on $\GS{m}$.
Indeed, we can always write it in its diagonal form $\tilde{\phi}=\sum_{\lambda}\lambda \ketbra{\lambda}$.
Then using the linearity of $e^{t\mathcal L}$ and the previous result on pure states, it is clear that $e^{t\mathcal L}(\tilde{\phi})$ is approximately supported on $\GS{m}$.
Second, let us remark that we are not proving in general that any state supported on $\GS{m}$ will evolve for $t\to\infty$ to a state in $\GS{n}$.
What we are showing is that at a certain time $t\gtrsim\poly\log N$ the state $e^{t\mathcal L}(\tilde{\phi})$ is close to a state in $\GS{n}$.
The evolution after such a time $t$ is not controlled, since in that time regime Lieb-Robinson bounds are not very strict.
This will prove to be enough for our purposes.
However notice that if $G(\ell)$ is purely exponential, namely $G(\ell)\propto e^{vt-\gamma\ell}$, then at a time $t=\alpha\ell$ for some $0<\alpha<1$, we can still follow our argument by choosing $\ell=\beta L$, with $0<\alpha<\beta<1$.
In this case one can still freely move the logical operators away from the fattening of the support of $h_i^{(n)}$.
Since the result is true for a time which is proportional to the linear size of the system $L$, then one may expect to be able to prove that in the thermodynamic limit, at infinite time $\tilde{\phi}$ will evolve to some state $\tilde{\psi}$ supported in $\GS{n}$.
However, this is beyond the scope of this work.

\paragraph{Multiplicative properties of the fattened logical operators.}
The main ingredient of the proof is the set of vectors obtained by applying the fattening of the logical operators to the initial vector $\ket{\phi}$, namely $\ket{\phi_{\alpha,\beta}} = \fat[\ell]{\lX_x^\alpha}\fat[\ell]{\lX_y^\beta}\ket{\phi}$ for $\alpha,\beta=0,\dots,n-1$.
In the following we will show that under the hypothesis, these vectors are close to vectors in $\GS{m}$ and their overlap matrix has full rank.
This will be the contradiction we are looking for, since $n>m$.
The fundamental step of the proof is about the multiplicative properties of the fattening of the logical operators.

\begin{step}
For any operator in the ground state space of the $D(\mathbb Z_m)$ model, $\Theta\in\mathcal B(\GS{m})$,
\begin{equation}
\Tr[\Theta\,\fat[\ell]{\lX_{x,y}^\alpha}\fat[\ell]{\lX_{x,y}^\beta}] = \Tr[\Theta\,\fat[\ell]{\lX_{x,y}^\alpha\lX_{x,y}^\beta}] + \delta_N,
\end{equation}
where the error $\delta_N$ vanishes in the thermodynamic limit.
\end{step}

This result is a direct consequence of the Schwarz inequality for completely positive and trace preserving maps, and we will prove it by following standard techniques used to generalize or specialize this inequality.
The first step consists in showing that the position $x$ or $y$ of any fattened operator $\fat[\ell]{\lX_{x,y}^\alpha}$ is irrelevant when it is applied to any vector $\ket*{\tilde\phi}\in\GS{m}$.
In the following we will focus on the operators defined on a vertical strip $\fat[\ell]{\lX_{x}^\alpha}$, the case for the horizontal operators $\fat[\ell]{\lX_{x}^\alpha}$ is completely analogous.
In other words we want to prove that $\fat[\ell]{\lX_x^\alpha}\ket*{\tilde\phi} \simeq \fat[\ell]{\lX_{x'}^\alpha}\ket*{\tilde\phi}$.
To do so, let us first show that they have norm smaller or equal than one, and that their overlap is one.
The first claim is easily checked using the fact that any Lindbladian evolution is norm-1 contractive, $\norm*{e^{t\mathcal L}(A)}_1 \leq \norm{A}_1$, and by duality its dual is norm-infinity contractive, $\norm*{e^{t\mathcal L^\star}(A)}_\infty \leq \norm{A}_\infty$.
Since $\norm*{\lX_x^\alpha}_\infty = 1$, then $\norm*{\fat[\ell]{\lX_x^\alpha}}_\infty \leq 1$.
Finally, for any operator $A$ with $\norm{A}_\infty\leq 1$, the norm of the vector $A\ket{\phi}$ for any normalized $\ket{\phi}$, is less or equal than one. 
The second claim can also be easily seen by choosing $x'$ so that $d(x,x')>2\ell$
\begin{equation}
\expval*{\fat[\ell]{\lX_x^{-\alpha}}\fat[\ell]{\lX_{x'}^\alpha}}{\tilde{\phi}} = \expval*{\fat[\ell]{\lX_x^{-\alpha}\lX_{x'}^\alpha}}{\tilde{\phi}}
= \expval*{e^{t\mathcal L}\big(\lX_x^{-\alpha}\lX_{x'}^\alpha\big)}{\tilde{\phi}} + \err{\ell L\, G(\ell)}.
\end{equation}
In the first passage we used the fact that the two fattened operators have disjoint supports on the lattice so trivially the product of the fattening is the fattening of the product.
In the second equality, we used again Lieb-Robinson bound.
The factor $\ell L$ in the error comes from the size of the support of the logical operators.
We also used the fact that any Lindbladian evolution preserves the hermitian conjugation, $\big(e^{t\mathcal L}(A)\big)^\dag = e^{t\mathcal L}(A^\dag)$.
To conclude, we now use duality, the previous result on the evolution of any vector in $\GS{m}$, and the fact that the $\lX_x$ are logical operators in $\GS{n}$ and therefore their position is immaterial on $\tilde{\psi}$, 
\begin{equation}
\label{eq:moving X}
\expval*{\fat[\ell]{\lX_x^{-\alpha}}\fat[\ell]{\lX_{x'}^\alpha}}{\tilde{\phi}}
= \Tr[\lX_x^{-\alpha}\lX_{x'}^\alpha\,\tilde{\psi}] + \err{LG(\ell),\eps_N} = 1 + \err{\ell L\,G(\ell),\eps_N}.
\end{equation}
Since the overlap of the two vectors is approximately one, we have also shown that they are approximately normalized.
Given eq.~\eqref{eq:moving X}, it is immediate to see that for any state $\rho$ supported on $\GS{m}$,
\begin{equation}
\label{eq:SI hyp}
\Tr[\rho\,e^{t\mathcal L^\ast}\big(\lX_x^{-\alpha}\big)e^{t\mathcal L^\ast}\big(\lX_x^{\alpha}\big)]\simeq
\Tr[\rho\,e^{t\mathcal L^\ast}\big(\lX_x^{-\alpha}\lX_x^{\alpha}\big)]\simeq 1.
\end{equation}
One just has to follow the same steps as before, knowing that the positions $x$ and $x'$ of both operators can be freely moved when they multiply $\rho$.
From here on, we will not write down explicitly the error:
as in previous sections `$\simeq$' will mean that all equalities have to be considered up to errors that vanish fast in the thermodynamic limit. In this section they will be of the same order of the ones discussed before, for example in eq.~\eqref{eq:moving X}, coming from Lieb-Robinson bounds and the fast evolution of $\mathcal L$. They vanish in the thermodynamic limit faster than any power in $N$.

We are now finally ready to apply Schwarz inequality.
For a completely positive and trace preserving map $T$, Schwarz inequality reads $T^\ast(A^\dag)T^\ast(A)\leq T^\ast(A^\dag A)$ for any matrix $A$.
It is easily proven in this setting by using Stinespring's representation.
It is also well known that if the equality holds for a matrix $A$, then $A$ belongs to the right multiplicative domain, namely $T^\ast(B)T^\ast(A)=T^\ast(BA)$ for any matrix $B$.
On the other hand, if the equality holds for $A^\dag$, then $A$ belongs to the left multiplicative domains, $T^\ast(A)T^\ast(B)=T^\ast(AB)$ for any matrix $B$.
This fact is easily checked by writing the Schwarz inequality for $A'=tA+B$ and $A''=tA+\iu B$.
The terms proportional to $t^2$ cancel out by the hypothesis on $A$, and for the inequality to be true for any $t\in\mathbb{R}$, the term proportional to $t$ must  vanish too.
The result is the desired multiplicative property.
In our case, the equality for the operator $e^{t\mathcal L^\ast}\big(\lX_x^\alpha\big)$ is not true in general, however we proved eq.~\eqref{eq:moving X} for any $\rho$ in $\GS{m}$.
By following the proof just sketched with this different hypothesis, one can easily see that 
$\Tr[\rho\,e^{t\mathcal L^\ast}\big(\lX_x^{\alpha}\big)e^{t\mathcal L^\ast}\left(B\right)]\simeq \Tr[\rho\,e^{t\mathcal L^\ast}\big(\lX_x^{\alpha}\,B\big)]$, and $\Tr[\rho\,e^{t\mathcal L^\ast}\left(B\right)e^{t\mathcal L^\ast}\big(\lX_x^{\alpha}\big)]\simeq \Tr[\rho\,e^{t\mathcal L^\ast}\big(B\,\lX_x^{\alpha}\big)]$.
Lastly, we notice that any operator $\Theta$ can be written as a linear combination of positive semidefinite operators.
For example one can write $\Theta$ as the sum of its hermitian and anti-hermitian part, and in turn write them as the sum of a positive minus a negative part.
Therefore the previous results are valid for any operator $\Theta\in\mathcal B(\GS{m})$,
\begin{subequations}
\label{eq:SI conclusion}
\begin{align}
&\Tr[\Theta\,e^{t\mathcal L^\ast}\big(\lX_x^{\alpha}\big)e^{t\mathcal L^\ast}\big(B\big)]\simeq 
 \Tr[\Theta\,e^{t\mathcal L^\ast}\big(\lX_x^{\alpha}\,B\big)], \\
&\Tr[\Theta\,e^{t\mathcal L^\ast}\big(B\big)e^{t\mathcal L^\ast}\big(\lX_x^{\alpha}\big)]\simeq 
 \Tr[\Theta\,e^{t\mathcal L^\ast}\big(B\,\lX_x^{\alpha}\big)].
\end{align}
\end{subequations}

\paragraph{A basis of $\GS{m}$ built from the fattened logical operators.}
With the previous results we are finally ready to show the desired contradiction that will conclude this section.

\begin{finalstep}
The matrix $T=\{\braket{\phi_{\alpha,\beta}}{\phi_{\gamma,\delta}}\}_{(\alpha,\beta),(\gamma,\delta)}$ with elements the overlaps of the vectors $\ket{\phi_{\alpha,\beta}}=\fat[\ell]{\lX_x^\alpha}\fat[\ell]{\lX_y^\beta}\ket{\phi}$ for $\alpha,\beta=0,\dots,n-1$ has determinant $\abs{\det T}\geq \delta >0$ uniformly in the system size $N$.
Therefore, for any $N$ and in the thermodynamic limit the vectors are linearly independent.
Moreover, they are $\eps$-close to states in $\GS{m}$, with $\eps$ vanishing in the thermodynamic limit.
This is in contradiction with the hypothesis, since $\dim[\GS{m}]=m^2$ on the torus and $m<n$.
\end{finalstep}

First, let us show that the $\ket{\phi_{\alpha,\beta}}$ are close to the ground state space $\GS{m}$.
We already showed that the horizontal operator can (approximately) be moved freely when acting on $\ket{\phi}$ and this fact immediately implies that the (approximately) normalized vector $\ket{\phi_{0,\beta}}=\fat[\ell]{\lX_y^\beta}\ket{\phi}$ are $\eps$-close to vectors in the ground state space of the $D(\mathbb{Z}_m)$ model, with vanishing $\eps$ in the thermodynamic limit.
To prove this claim consider any local, frustration free Hamiltonian $H^{(m)}=\sum_i h_i^{(m)}$ whose ground state space is $\GS{m}$.
Then $H^{(m)}\ket{\phi_{0,\beta}} = \sum_ih_i^{(m)}\ket{\phi_{0,\beta}}$ and $h_i^{(m)}\ket{\phi_{0,\beta}}= h_i^{(m)} \fat[\ell]{\lX_y^\beta} \ket{\phi}$.
Thanks to the properties of the `fattened' operators, we can choose $\fat[\ell]{\lX_y^\beta}$ to have a support with empty intersection with the support of $h_i^{(m)}$ (notice that by changing the support we pick up errors that go to zero in the thermodynamic limit).
Then we can conclude $h_i^{(m)} \ket{\phi_{0,\beta}}\simeq\fat[\ell]{\lX_y^\beta} h_i^{(m)}\ket{\phi}=0$.
The state $\ket{\phi_{\alpha,\beta}}$ can be written as $\ket{\phi_{\alpha,\beta}}=\fat[\ell]{\lX_x^\alpha}\ket{\phi_{0,\beta}}$,
and using again eq.~\eqref{eq:moving X} we conclude that also the horizontal coordinate $x$ can be chosen at will (again up to errors vanishing in the thermodynamic limit) since $\fat[\ell]{\lX_x^\alpha}$ is acting on an approximate ground state.
Using the same argument as before, for any Hamiltonian term $h_i^{(m)}$ we can choose the position of the fattened operators such that their support has no overlap with the support of $h_i^{(m)}$ and we conclude that $\ket{\phi_{\alpha,\beta}}$ is $\eps$-close to $\GS{m}$, with $\eps\to 0$ in the thermodynamic limit.

The last step consists in showing that the overlap matrix $T=\{\braket{\phi_{\alpha,\beta}}{\phi_{\gamma,\delta}}\}_{(\alpha,\beta),(\gamma,\delta)}$ has full rank.
For this step it is crucial the multiplicative property in eq.~\eqref{eq:SI conclusion}.
Consider the overlap
$\braket{\phi_{\alpha,\beta}}{\phi_{\gamma,\delta}} = \matrixel{\phi_{\alpha,\beta}}{\fat[\ell]{\lX_x^{\gamma}}\fat[\ell]{\lX_y^{\delta}}}{\phi}$.
Using eq.~\eqref{eq:SI conclusion} with $\Theta=\ketbra{\phi}{\phi_{\alpha,\beta}}$ we have 
$\braket{\phi_{\alpha,\beta}}{\phi_{\gamma,\delta}} \simeq \matrixel{\phi_{\alpha,\beta}}{\fat[\ell]{\lX_x^{\gamma}\lX_y^{\delta}}}{\phi}\simeq \matrixel{\phi_{0,\beta}}{\fat[\ell]{\lX_x^{-\alpha}}\fat[\ell]{\lX_x^{\gamma}\lX_y^{\delta}}}{\phi}$.
It is now clear how to proceed:
We use again eq.~\eqref{eq:SI conclusion} with $\Theta=\ketbra{\phi}{\phi_{0,\beta}}$ and subsequently with $\Theta=\ketbra{\phi}$, to obtain 
$\braket{\phi_{\alpha,\beta}}{\phi_{\gamma,\delta}}\simeq\expval{\fat[\ell]{\lX_y^{-\beta}\lX_x^{-\alpha}\lX_x^{\gamma}\lX_y^{\delta}}}{\phi}$.
Finally, we use Lieb-Robinson bounds, duality, and the hypothesis of the fast evolution  $\phi\fastto[\mathcal L]\psi$ to conclude that
$\braket{\phi_{\alpha,\beta}}{\phi_{\gamma,\delta}}=\braket{\psi_{\alpha,\beta}}{\psi_{\gamma,\delta}}+\err{\eps_N,\ell L\,G(\ell)}$.
For a large enough system size $N$, the error term can be made small enough such that $\abs{\det S_N - \det T_N}< 1/2\abs{\det S_N}$.
Since the determinant of $S_N$ is strictly different from zero for any $N$, then $\abs{\det T}>1/2\abs{\det S}$, and the overlap matrix has full rank uniformly in $N$.

This concludes the no go theorem, since we found a set of vectors which generate a subspace of dimension $n^2$ for any $N$, and that are all $\eps$-close to $\GS{m}$ with $\eps\to 0$ in the $N\to\infty$ limit.
This contradicts the fact that $\dim[\GS{m}]=m^2$ and $m<n$.

Before closing this section let us make some observations.
First of all, it should be clear from the exposition that the strategy and all the arguments used can be easily generalized to the case of higher genus.
We believe that the general case would lead only to a more involved notation and exposition, without adding anything to the discussion.
For this reason we decided to present only the  case of genus $g=1$.
Moreover, the result can also be generalized to planar codes with boundaries.
Actually, in the easiest case of a square geometry with $x$-boundaries in the horizontal direction and $z$-boundaries in the vertical direction (for definitions, see Ref.~\cite{Bravyi1998}), all the proofs are even simpler, since one has to deal only with either vertical $\lX$ operators or horizontal $\lZ$ operators.
Therefore, the multiplicative property eq.~\eqref{eq:SI conclusion} is not needed.

Second, it would be interesting to address also the more general case in which the initial and the final state $\phi$ and $\psi$ are mixed and supported on the ground state space of respectively $D(\mathbb Z_m)$ and $D(\mathbb Z_n)$.
However, in this case we should drastically change the approach to the proof, since counting the number of ground states does not seem the appropriate strategy in the mixed case.

Finally, the last comment is about the case of genus zero, as for example the case of a system living on a sphere.
In this case our approach fails since there are no logical operators and the ground state is unique.
The underlying topological properties of such state can be studied in many ways.
For example, in Ref.~{\cite{Haah2014}} the author considers certain sets of operators living on rings.
These operators would give rise to logical operators in a geometry with genus greater than zero, when they revolve around the fundamental cycles.
They can be viewed as the physical process of creating a pair of anyons, bringing them far apart, and subsequently annihilate each other.
With such operators defined on large rings, it is possible to define a set of projectors that detect the presence of anyonic excitation in the interior of the ring.
In this sense, they are identified with the particle superselection sectors.
Finally, the author builds the topological $S$-matrix by linking pairs of these ring operators and computing the ground state expectation values of the resulting operator.
This object is proven to depend only on the state under study, and not on any Hamiltonian with such as a ground state.
It would be interesting to study what is the effect of a fast dissipative evolution onto these objects.
Following this line of thoughts it may be possible to prove the result of this Section also in the case of genus zero.
However, this task contains many subtleties and many non trivial intermediate steps, and is therefore left for future work.

\subsection{Fast dissipative evolution between 1D `topological states'}
\label{subsec:anyon condensation}

In order to better understand the implications of the definitions of phases through fast dissipative dynamics it would be nice to find an example of Lindbladian $\mathcal L$ which drives in short time a state in a certain phase to another one which belongs to a different phase whose long range correlations are contained in the original one.
More concretely, as an example we would like to show the possibility of performing in short time the inverse process with respect to Section~\ref{subsec:slow Zm to Zn}.
We expect it would be possible to take any state $\ket{\psi}\in\GS{n}$ and devise a dissipative evolution which drives it in short time to a state $\ket{\phi}\in\GS{m}$, with $m$ a divisor of $n$.
In this case, indeed the anyon content of the $D(\mathbb Z_m)$ model is a subset of the anyon content of the $D(\mathbb Z_n)$ model.
More generally, we expect this process to be possible for any topological model described by a group $G$ with a normal subgroup $H$.

Such a fast dissipative dynamics would be very interesting also since it would provide a physical dynamic implementation of the mechanisms of anyon condensation and confinement~\cite{Bais2009}.
Anyon condensation is the process where an anyonic excitation of the original model stops describing an excitation and becomes part of the vacuum of the final model.
On the other hand, we say that an anyon excitation is confined when the energy of the excitation depends on the distance between the two anyons (they are always created in pairs).
In practice the flux connecting the two anyons cannot be moved freely, hence the excitation is not topological any more.
In particular, if we take the two anyons very far apart, the energy of the resulting state is infinite, therefore the state is not physical any more (its norm vanishes).
This also means that we cannot have an isolated anyonic excitation, since its outgoing flux would have infinite energy.
The situation is equivalent to creating a pair of anyons and then bring one of them to infinity.
The two mechanisms of anyon condensation and confinement are not independent, on the contrary they are dual to each others.
When an anyon is condensed, the anyons that braid non-trivially with it are confined.

Recently anyon condensation and confinement have been studied with PEPS with both analytical and numerical tools~\cite{Duivenvoorden2017,Iqbal2017a,Garre-Rubio2017}.
It has been shown that changing the original PEPS tensor, either smoothly or abruptly, it is possible to implement anyon condensation and confinement:
some topological excitations become localized while others are reduced to the trivial sector.
We conjecture that it should be possible to implement the same transformation dynamically, with a fast Lindbladian evolution.
Notice that the results of Section~\ref{subsec:slow Zm to Zn} are based on the fact that with a local Lindbladian it is not possible to generate in short times long range correlations that were not present in the initial state.
However we expect that it should be possible to selectively destroy some of the long range correlations that are present in the initial state.
The most trivial example of this fact is that it is always possible to evolve in short times any state to the product state (for example using the Lindbladian described in Section~\ref{subsec:1D SPT phase}), where no correlations of any kind are present.
Our claim is stronger:
We think indeed it should be possible to destroy only a specific subset of the anyon content of a topological state, while leaving the rest untouched.

This task is much more complicated to achieve than the case of the final product state, since the Lindbladian must be very fine tuned.
It is pretty involved to devise a Lindbladian that implements in short time anyon condensation for topological states in two dimensions, even in the simplest case of $D(\mathbb Z_4)$ condensing to $D(\mathbb Z_2)$, which has been studied in great detail in Ref.~\cite{Iqbal2017a}.
While we leave this for future work, we can give a much simpler example of this process in one dimension.

Consider a one dimensional lattice $\Lambda$ of size $N$ with periodic boundary conditions.
The analogous of $D(\mathbb Z_n)$ is the GHZ state, $\ket{\text{GHZ}_n} \propto \sum_{\alpha=0}^{n-1}\ket{\alpha,\dots,\alpha}$.
GHZ states can be chosen as the representatives of phases in the absence of symmetry restrictions~\cite{Schuch2011}.
For these states, the parent Hamiltonian has a degenerate ground state space $\mathcal H_0(\text{GHZ}_n)$ spanned by the states $\ket{\psi_{\beta}}\propto\sum_{\alpha=0}^{n-1}\omega_n^{\alpha\beta}\ket{\alpha,\dots,\alpha}$ for $\beta=0,\dots,n-1$.
Notice that the product states $\ket{\alpha,\dots,\alpha}$ belong to the ground state space generated by the $\ket{\psi_\alpha}$.
This does not mean that the GHZ states are in the trivial phase as the product state.
Indeed, by looking at correlation functions, it is not difficult to show that there cannot exist any finite depth quantum circuit mapping a product state to a GHZ state. 

Moreover, the results of the Section~\ref{subsec:slow Zm to Zn} apply here:
There does not exist any fast Lindbladian evolution such that $\text{GHZ}_m\fastto[\mathcal L]\text{GHZ}_n$ if $m<n$.
The way to prove this claim is exactly the same as before.
The states $\ket{\psi_{\beta}}$ can be obtained from $\ket{\psi_0}$ by applying a single-site local operator $\ket{\psi_{\beta}}=Z_x^\beta\ket{\psi_0}$, where the site $x$ can be chosen anywhere in the ring.
Then, by contradiction, if it exists a dissipative evolution that drives fast $\text{GHZ}_m\fastto[\mathcal L]\text{GHZ}_n$ we define the fattened operators $\fat[\ell]{Z_x^\beta} = e^{t\mathcal L_{B_{x,\ell}}}(Z_x^\beta)$ and a set of states $\ket{\phi_\beta}=\fat[\ell]{Z_x^\beta} \ket{\text{GHZ}_m}$ for $\beta=0,\dots,n-1$.
Analogously to Section~\ref{subsec:slow Zm to Zn}, the crucial property is that $\fat[\ell]{Z_x^\beta} \ket{\phi_0} \simeq \fat[\ell]{Z_{x'}^\beta} \ket{\phi_0}$.
This can be shown by taking $x$ and $x'$ far apart, such that $d(x,x')>2\ell$, and computing the overlap $\expval{\fat[\ell]{Z_{x}^{-\beta}}\fat[\ell]{Z_{x'}^\beta}}{\phi_0}$.
In this case $\fat[\ell]{Z_{x}^{-\beta}}\fat[\ell]{Z_{x'}^\beta}=\fat[\ell]{Z_{x}^{-\beta}Z_{x'}^\beta}$ and by using Lieb-Robinson bounds, the hypothesis of fast evolution and the fact that the position $x$ is immaterial when $Z_x$ is applied on $\ket{\psi_0}$, it is easy to show that the overlap is approximately one, with errors that vanish in the thermodynamic limit.
With this result, it is easy to show that the overlap matrix $\braket{\phi_{\alpha}}{\phi_\beta}\simeq \braket{\psi_\alpha}{\psi_\beta}$ and therefore for some large enough system the states $\ket{\phi_\beta}$ for $\beta=0,\dots,n-1$ are linearly independent.
Moreover, since the operator $\fat[\ell]{Z_x^\beta}$ can be moved freely when acting on $\ket{\phi_0}$ (up to errors vanishing in the thermodynamic limit), it is easy to see that $\ket{\phi_\beta}$ is an approximate ground state of any frustration free local Hamiltonian whose ground state is $\mathcal H_0(\text{GHZ}_m)$.
This provides a contradiction, since $\mathcal H_0(\text{GHZ}_m)$ has dimension $m<n$.

However, we can easily prove that the opposite procedure can be performed when $m$ is a divisor of $n$, namely we can devise a fast Lindbladian evolution such that $\text{GHZ}_m\fastto[\mathcal L]\text{GHZ}_n$.
For the sake of simplicity, let us restrict to the case $n=4$ and $m=2$.
Consider the following CPTP local map acting on site $i$
\begin{equation}
\mathcal T_i(\rho) = P_i\rho\,P_i + X_i^\dag(1-P_i)\,\rho\,(1-P_i) X_i,
\end{equation}
where $P_i$ is the projector $P_i = \ketbrasub{0}{0}{i} + \ketbrasub{2}{2}{i}$ and $X_i$ is a unitary ladder operator $X_i = \sum_{\alpha=0}^{n-1}\ketbrasub{\alpha+1}{\alpha}{i}$.
We then define the strictly local Lindbladian $\mathcal L_i=\mathcal T_i - \id$ and finally $\mathcal L = \sum_i \mathcal L_i$.
It is easy to check that the single site CPTP map is idempotent, $\mathcal T_i^2=\mathcal T_i$.
Then, using eq.~\eqref{eq:fast local L} and the discussion thereafter, we know that for any initial state, $\mathcal L$ converges fast, after a time $t\gtrsim \log N$, to the state $\mathcal T_\Lambda(\rho)$.
Finally, it is easy to check that $T_\Lambda(\text{GHZ}_4) = \text{GHZ}_2$ as desired.

Notice that due to the nature of dissipative dynamics, at any finite time $t$ the evolved state is mixed.
However, the evolution is fine tuned such that  the final state is the desired pure state $\ket{\text{GHZ}_2}$.
For more complicated cases, like two dimensional topological states, it is not easy in general to devise Lindbladian whose final state is pure.
On a side, this raises the question whether it is possible to define topological order for generic mixed states.
The task has already been considered in the case of Gibbs states in Ref.~\cite{Iblisdir2009,Hastings2011}.
A more general approach is desirable, however it goes beyond the scopes of this work and is left for future work.

\section{Conclusions and outlook}

In this work, motivated by~\cite{Konig2014}, we introduced a new definition of quantum phases valid for mixed states.
Roughly speaking, the standard definition of quantum phases for pure states establish that two states are in the same phase if there is a finite depth quantum circuit that drives one state to the other.
In the new definition we allow the circuit to be composed of quantum channels instead of unitaries, which makes it irreversible.
Instead of just sticking to such definition we impose that the noisy circuit corresponds to the time evolution of a time-independent local Lindbladian.
Besides the obvious fact that the new allowed evolutions are simpler, 
it has the additional advantage of giving a continuous path between the two states on which observables behave smoothly.
It gives an operational point of view to the definition that connects with the usual notion of phase and phase transition as the absence or presence of divergences in the behaviour of observables.

In this work we give the first steps that show that the proposed definition is a reasonable choice.
First of all, it divides all mixed state into equivalent classes, that are further partially ordered according to their topological complexity.

We also show that, when it comes to standard topological order, for pure states, the new definition gives exactly the same equivalent classes.
One implication is proven in full generality, the other only for the $\mathbb{Z}_n$ quantum double phases in two dimensions.
However, in the case of symmetry protected phases in 1D, all states belong to the trivial phase with the Lindbladian definition, something that separates substantially from the Hamiltonian case. 

Being the first work in which a definition is proposed, there are many interesting questions that are left open for future work.
Among them, arguably the most relevant one is how topological phases are ordered according to the new definition.
Some very preliminary results made us conjecture that if a given topological state can be obtained from another via a process of anyon condensation, then there is a fast Lindbladian that can also drive the first state into the second.
This would make anyon condensation the mechanism that orders the phases according to the definition that we propose here.

\begin{acknowledgements}
We are grateful to F.~Brand{\~a}o, B.~Nachtergaele, N.~Schuch, J.~Garre-Rubio and S.~Iblisdir for helpful conversations and correspondence.
This project has received funding from the European Research Council (ERC) under the European 15 Union’s Horizon 2020 research and innovation programme through the ERC Consolidator Grant GAPS (No. 648913).
We acknowledge financial support from MINECO (grant MTM2014-54240-P, MTM2017-88385-P and Severo Ochoa project SEV-2015-556) and Comunidad de Madrid (grant QUITEMAD+-CM, ref. S2013/ICE-2801).
\end{acknowledgements}

\printbibliography 

\end{document}